\documentstyle[12pt,epsf]{article}
\def\photonatomright{\begin{picture}(3,1.5)(0,0)
                                \put(0,-0.75){\tencircw \symbol{2}}
                                \put(1.5,-0.75){\tencircw \symbol{1}}
                                \put(1.5,0.75){\tencircw \symbol{3}}
                                \put(3,0.75){\tencircw \symbol{0}}
                      \end{picture}
                     }

\def\photonatomup{\begin{picture}(1.5,3)(0,0)
                             \put(-0.75,3){\tencircw \symbol{3}}
                             \put(-0.75,1.5){\tencircw \symbol{2}}
                             \put(0.75,1.5){\tencircw \symbol{0}}
                             \put(0.75,0){\tencircw \symbol{1}}
                   \end{picture}
                  }

\def\photonright{\begin{picture}(30,1.5)(0,0)
                     \multiput(0,0)(3,0){10}{\photonatomright}
                  \end{picture}
                 }

\def\photonrighthalf{\begin{picture}(30,1.5)(0,0)
                     \multiput(0,0)(3,0){5}{\photonatomright}
                  \end{picture}
                 }

\def\photonuphalf{\begin{picture}(1.5,15)(0,0)
                      \multiput(0,0)(0,3){5}{\photonatomup}
                   \end{picture}
                  }

\def\fermionup{\begin{picture}(1,30)(0,0)
                     \put(0,0){\vector(0,1){15}}
                     \put(0,15){\line(0,1){15}}
               \end{picture}
              }
\def\fermionuphalf{\begin{picture}(1,15)(0,0)
                         \put(0,0){\vector(0,1){7.5}}
                         \put(0,7.5){\line(0,1){7.5}}
                   \end{picture}
                  }

\def\fermionull{\begin{picture}(30,15)(0,0)
                        \put(0,0){\vector(-2,1){15}}
                        \put(-15,7.5){\line(-2,1){15}}
                  \end{picture}
                 }
\def\fermionullhalf{\begin{picture}(15,7.5)(0,0)
                        \put(0,0){\vector(-2,1){7.5}}
                        \put(-7.5,3.75){\line(-2,1){7.5}}
                  \end{picture}
                 }
\def\fermionurr{\begin{picture}(30,15)(0,0)
                        \put(-30,-15){\vector(2,1){15}}
                        \put(-15,-7.5){\line(2,1){15}}
                  \end{picture}
                 }
\def\fermionurrhalf{\begin{picture}(15,7.5)(0,0)
                        \put(-15,-7.5){\vector(2,1){7.5}}
                        \put(-7.5,-3.75){\line(2,1){7.5}}
                  \end{picture}
                 }

\newenvironment{Feynman}[3]{\begin{center}
                            \setlength{\unitlength}{#3 mm}
                            \begin{picture}(#1)(#2)
                            \thicklines
                           }{\end{picture} \end{center}}
\def\theequation{\arabic{section}.\arabic{equation}}
\newcommand{\ezero}{\setcounter{equation}{0}}

\newcommand{\nn}{\noindent}

\newcommand{\bq}{\begin{equation}}
\newcommand{\eq}{\end{equation}}
\newcommand{\ba}{\begin{eqnarray}}
\newcommand{\ea}{\end{eqnarray}}

\newcommand{\mr}{\mathrm}
\newcommand{\ww}{\Gamma_W}
\newcommand{\wm}{M_W}
\newcommand{\nl}{ \nonumber \\}
\newcommand{\nll}{ \nonumber \\}
\begin{document}
\thispagestyle{empty}
\onecolumn
\begin{flushleft}
{DESY 95--167\\}
{hep-ph/9509abc\\}
September 1995
\end{flushleft}
\vspace*{3.0cm}
\nn
\hspace*{-5mm}
\begin{center}
{\LARGE
Off shell $W$ pair production in $e^+ e^-$ annihilation:

\bigskip

The {\tt CC11} process
}

\vspace*{1.0cm}

{\large
D. Bardin$\,^{1,2}$ and $\;$ T.~Riemann$\,^{1}$
\\
\vspace*{1.0cm}
}
\end{center}

\begin{normalsize}
\begin{tabbing}
$^1$ \=
DESY -- Zeuthen
\\ \>
  Platanenallee 6, D-15738 Zeuthen, Germany
\end{tabbing}
\begin{tabbing}
$^2$  \=
Bogoliubov Laboratory for Theoretical Physics, JINR
\\ \>
ul. Joliot-Curie 6, RU-141980 Dubna, Moscow Region, Russia
\end{tabbing}
\end{normalsize}

\vspace*{1.5cm}
\date{\today}
\vfill
\thispagestyle{empty}
{\large
\centerline{
ABSTRACT}
\vspace*{.3cm}
\nn
\normalsize
The various four-fermion production channels in $e^+ e^-$ annihilation
are discussed and the {\tt CC11} process $e^+ e^- \rightarrow {\bar
  f}_1^u f_1^d f^u_2 {\bar f}_2^d$, {\mbox{$f$$\neq$$e$}}, is studied
in detail. The cross section $d^2\sigma/ds_1ds_2$, with $s_1,s_2$
being the invariant masses squared of the two fermion pairs, may be expressed
by six generic functions. All but one may be found in the literature.
The cross section, including initial state radiation and the Coulomb
correction, is discussed and compared with other calculations from low
energies up to $\sqrt{s}$ =  2 ~TeV.
\vspace*{.3cm}
\nn
\normalsize
} 
\vspace*{.5cm}

\bigskip
\vfill
\footnoterule
\nn
{\small
\begin{tabbing}
Email addresses: \= bardindy@cernvm.cern.ch, riemann@ifh.de
\end{tabbing}
}
\newpage
%
\section{Introduction}
%
\ezero
A measurement of $W$ pair production
around and above the production threshold will be one of the main
tasks of LEP~2~\cite{lepproc}  as well as of a future
high energy linear $e^+ e^-$ collider~\cite{lcproc}.

\begin{figure}
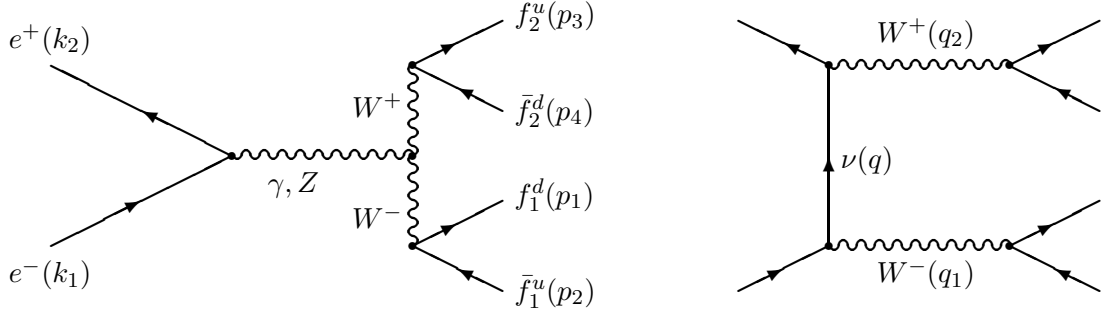

\begin{minipage}[tbh]{7.8cm}{
\begin{center}
\begin{Feynman}{75,60}{0,0}{0.8}
%
\put(30,30){\fermionurr}
\put(30,30){\fermionull}
\put(30,30){\photonright}
\put(30,30){\circle*{1.5}}
\put(60,30){\circle*{1.5}}
\put(60,15){\circle*{1.5}}
\put(60,45){\circle*{1.5}}
\put(75,7.5){\fermionullhalf}
\put(75,22.5){\fermionurrhalf}
\put(60,15){\photonuphalf}
\put(60,30){\photonuphalf}
\put(75,37.5){\fermionullhalf}
\put(75,52.5){\fermionurrhalf}
\small
\put(-07,48){$e^+(k_2)$}  
\put(-07,09){$e^-(k_1)$}
\put(36,24){$\gamma, Z$}
\put(50,36.5){$W^+$}  
\put(50,18.0){$W^-$}
\put(77,06){${\bar f}_1^u(p_2)$}  
\put(77,22){$       f_1^d(p_1)$}
\put(77,36){${\bar f}_2^d(p_4)$}  
\put(77,52){$       f_2^u(p_3)$}
\normalsize
\end{Feynman}
\end{center}
}\end{minipage}
\begin{minipage}[tbh]{7.8cm} {
\begin{center}
\begin{Feynman}{75,60}{0,0}{0.8}
%
\put(30,15){\fermionurrhalf}
\put(30,45){\fermionullhalf}
\put(30,15){\fermionup}
\put(30,45){\photonright}
\put(30,15){\photonright}
\put(30,45){\circle*{1.5}}
\put(30,15){\circle*{1.5}}
\put(60,15){\circle*{1.5}}
\put(60,45){\circle*{1.5}}
\put(75,7.5){\fermionullhalf}
\put(75,22.5){\fermionurrhalf}
\put(75,37.5){\fermionullhalf}
\put(75,52.5){\fermionurrhalf}
\small
\put(32,28){$\nu(q)$}
\put(38,49){$W^+(q_2)$}  
\put(38,09){$W^-(q_1)$}
\normalsize
\end{Feynman}
\end{center}
}\end{minipage}
\vspace*{.5cm}

\noindent
\caption{
\label{fig.1}
\it
The double resonating {\tt CC3} contributions to off shell $W$ pair production:
{\tt crayfish} and {\tt crab}.
They are part of all the processes of table~1.
}
\vspace*{.5cm}
\end{figure}

Immediately after their creation, the $W$ bosons decay and
four-fermion ($4f$) production is observed:
\ba
e^+ e^- \rightarrow (W^- W^+) \rightarrow 4 f.
\label{eqborn}
\ea
This double resonating process proceeds via the diagrams shown in
figure~\ref{fig.1}.
In addition, there are a lot of Feynman diagrams with the same final state,
but different intermediate states, which are single resonant or
non-resonant and often called
background diagrams.
Their number and complexity
vary in dependence on the composition of the final state.
The total numbers of Feynman diagrams for the different channels are
shown in table~1.

\begin{table}[hbt]
\label{tab1}
\begin{center}
\begin{tabular}{|c|c|c|c|c|c|}
\hline
             &
\raisebox{0.pt}[2.5ex][0.0ex]{${\bar d} u$}
& ${\bar s} c$ & ${\bar e} \nu_{e}$ &
              ${\bar \mu} \nu_{\mu}$ & ${\bar \tau} \nu_{\tau}$   \\
\hline
$d {\bar u}$            &{\it  43}& {\bf 11} &  20 & {\bf 10} & {\bf 10} \\
\hline
$e {\bar \nu}_{e}$      &  20 &  20 &{\it 56}&  18 &  18 \\
\hline
 $\mu {\bar \nu}_{\mu}$ & {\bf 10} & {\bf 10} &  18 & {\it 19} & {\bf 9}  \\
\hline
\end{tabular}
\caption
{\it
Number of Feynman diagrams contributing to the production of two fermion
doublets.
\label{tab.1}
}
\end{center}
\end{table}

One may distinguish three different event classes, all of
them containing the {\tt CC3} process\footnote{
In~\cite{excalibur}, a slightly different classification has been introduced;
the relation of both schemes is discussed in~\cite{wwteup}.
}:
\begin{itemize}
\item[(i)]
    The {\tt CC11} process.
\\
    The two fermion pairs are different, the final state does not
    contain
    identical particles nor electrons or electron neutrinos
    (numbers in table~1 in {\bf boldface}). The corresponding eleven
    diagrams
    are shown in figures~1 and~2. There are less diagrams if neutrinos
    are produced    ({\tt   CC9, CC10} processes).

\item[(ii)]
    The {\tt CC20} process.
\\
    The final state contains one $e^{\pm}$ together
    with its neutrino (Roman numbers in table~1);
    compared to case~(i), the additional diagrams have a $t$ channel
    gauge boson exchange. For a purely leptonic final state, a {\tt CC18}
    process results.

\item[(iii)]
    The {\tt mix43} and {\tt mix56} processes.
\\
    Two mutually charge conjugated fermion pairs are produced ({\it
      italic} numbers in table~1).
    Differing from cases~(i) and~(ii), the diagrams may contain neutral
    boson exchanges\footnote
{
We exclude the Higgs boson exchange diagrams from the discussion.
In fact, the methods developed here are applicable also for the study
of
associated Higgs production~\cite{nc24h}.
}
    (see also table~2 in~\cite{nc24} where `neutral
    current' type final states are classified).
    There are less diagrams in the {\tt mix43} process if neutrinos
    are produced ({\tt mix19} process).
\end{itemize}

\begin{figure}
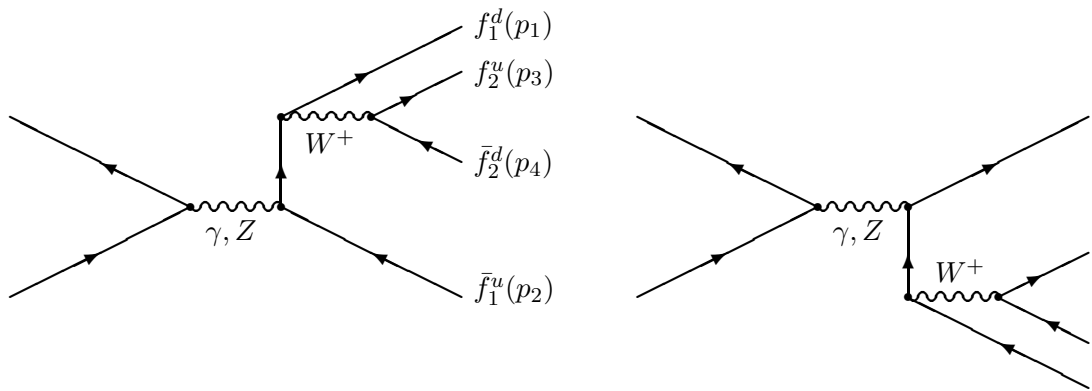

\vspace*{.3cm}
\begin{minipage}[tbh]{7.8cm}{
\begin{center}
\begin{Feynman}{75,60}{-5.0,0}{0.8}
%
\put(20,30){\fermionurr}
\put(20,30){\fermionull}
\put(20,30){\photonrighthalf}
\put(65,15){\fermionull}
\put(20,30){\circle*{1.5}}
\put(35,30){\circle*{1.5}}
\put(35,45){\circle*{1.5}}
\put(50,45){\circle*{1.5}}
\put(35,30){\fermionuphalf}
\put(65,60){\fermionurr}
\put(35,45){\photonrighthalf}
\put(65,52.5){\fermionurrhalf}
\put(65,37.5){\fermionullhalf}
\small
\put(22.5,24.5){$\gamma, Z$}
\put(39,39){$W^{+}$}
\put(67,59){$       f_1^d(p_1)$}
\put(67,51){${     f}_2^u(p_3)$}
\put(67,36){${\bar f}_2^d(p_4)$}  
\put(67,15){${\bar f}_1^u(p_2)$}  
\normalsize
\end{Feynman}
\end{center}
}\end{minipage}
\begin{minipage}[tbh]{7.8cm} {
\begin{center}
\begin{Feynman}{75,60}{0,0}{0.8}
%
\put(30,30){\fermionurr}
\put(30,30){\fermionull}
\put(30,30){\photonrighthalf}
\put(30,30){\circle*{1.5}}
\put(45,30){\circle*{1.5}}
\put(45,15){\circle*{1.5}}
\put(60,15){\circle*{1.5}}
\put(45,15){\fermionuphalf}
\put(75,00){\fermionull}
\put(75,45){\fermionurr}
\put(45,15){\photonrighthalf}
\put(75,22.5){\fermionurrhalf}
\put(75,07.5){\fermionullhalf}
\small
\put(32.5,24.5){$\gamma, Z$}
\put(49.5,17.5){$W^{+}$}
\normalsize
\end{Feynman}
\end{center}
}\end{minipage}
\vspace*{.5cm}

\noindent
\caption{\it
Single resonant contributions to off shell $W$ pair production:
a $u$-{\tt{deer}} and a $d$-{\tt{deer}}.
\label{fig.2}
}
\end{figure}

In this article, we will investigate the simplest topology, case~(i),
which proceeds via the diagrams of figures~\ref{fig.1} and~\ref{fig.2}:
\ba
e^+ e^- \rightarrow {\bar f}_1^u f_1^d f^u_2 {\bar f}_2^d,
\hspace{1cm}  f_1 \neq f_2, \, \, f_i \neq e.
\label{eqabs}
\ea

As far as {\it observable} final states are concerned, the following
reactions are covered by this study:
\begin{itemize}
\item[(i)]
(${\bar \mu} \nu_{\mu}$) + ($\tau {\bar \nu_{\tau}}$);
\item[(ii)]
($l {\bar \nu_{l}}$) + ($q_{u} {\bar q_{d}}$), \, $l=\mu,\tau$;
\item[(iii)]
(${\bar {u}} {d}$) + ($c {\bar s}$).
\end{itemize}

We parameterize the seven-dimensional four particle phase space as a
sequence of two particle phase spaces:
%
\begin{eqnarray}
\label{domega}
d\Gamma
&=& \prod_{i=1}^4\frac{d^3p_i}{2p_{i}^0}
 \times  \delta^4(k_1+k_2-\sum_{i=1}^4 p_i)
\nonumber
\\
&=&~\frac{\pi}{256}
\frac{\sqrt{\lambda(s,s_1,s_2)}}{s}
\frac{\sqrt{\lambda(s_1,m_1^2,m_2^2)}}{s_1}
\frac{\sqrt{\lambda(s_2,m_3^2,m_4^2)}}{s_2}
d s_1 \, d s_2\, d \cos\theta\,  d \Omega_1\, d \Omega_2,
\end{eqnarray}
with the usual definition of the $\lambda$ function,
\ba
\lambda(a,b,c) &=& a^2+b^2+c^2-2ab-2ac-2bc,
\\
\lambda &\equiv& \lambda(s,s_1,s_2).
\label{lambda}
\ea

\noindent
In~(\ref{domega}), the rotation angle around the beam axis has been
integrated over already.
Variables $k_1$ and $k_2$ are the four-momenta of
electron and positron and $p_1,p_2,p_3,p_4$ are those of the
final state particles $f_1^d,\bar{f}_1^u, f_2^u,\bar{f}_2^d$ with
$p_i^2=m_i^2~$\footnote{
Fermion masses are retained in the phase space definitions and in the
basic kinematical relations.
In all subsequent steps of the calculation, they
will be neglected compared to invariants $s, s_1$, and
$s_2$.}.
The invariant masses squared of the fermion pairs are:
\begin{eqnarray}
s&=&(k_1+k_2)^2,
\\
s_1&=&(p_1+p_2)^2,
\\
 s_2&=&(p_3+p_4)^2.
\end{eqnarray}
The  $W^-$ production angle $\theta$ in the center of mass system is
spanned by the vectors
($\vec{p}_1+\vec{p}_2$) and $\vec{k}_1$.
The spherical angle of $\vec{p_1}$
($\vec{p_3}$)
in the rest frame of the fermion pair [$f_1{\bar f}_1$]
([$f_2{\bar f}_2$]) is $\Omega_1\ (\Omega_2)$:
$d \Omega_i = d \cos\theta_i \, d \phi_i$.
In this frame, the $\vec{p_2}$ ($\vec{p_4}$) points into the opposite
direction with respect to $\vec{p_1}$ ($\vec{p_3}$) .
The kinematical ranges of the integration variables are:
\begin{eqnarray}
(m_1+m_2)^2     &\le&    s_1      \le (\sqrt{s}-m_3-m_4)^2,
\\
(m_3+m_4)^2     &\le&    s_2      \le (\sqrt{s}-\sqrt{s_1})^2 ,
\\
-1 &\le& \cos\theta,\ \cos\theta_1,\ \cos\theta_2   \le 1,
\\
0 &\le& \phi_1,\phi_2 \le 2\pi.
\end{eqnarray}

We are interested in analytical formulae for distributions in invariant masses
of fermion pairs.
Thus, we have to integrate analytically over the five angular
variables in (\ref{domega}).
The squared matrix elements have been derived with {\tt CompHEP}~\cite{comphep}
and the angular integrations were performed with the aid of {\tt
  FORM}~\cite{form}.

The cross section contributions may be grouped into several classes of
interferences:
\ba
\sigma_{\tt CC11}(s) =
\int ds_1 ds_2
\,
\left[ \sigma_{\tt CC3} + \sigma^{3f} + \sigma^{\nu f} + \sigma^{ff} +
\sigma^{f_1f_2} \right].
\label{tsig}
\ea
In~(\ref{tsig}),
the {\tt CC3} process is accompanied by the interferences  $\sigma^{3f}$
of {\tt deers} with the {\tt crayfish} diagrams,
the interferences $\sigma^{\nu f}$ of {\tt deers} with the {\tt crab},
the interferences  $\sigma^{ff}$ of {\tt deers} of one doublet,
and the interferences $\sigma^{f_1f_2}$ of {\tt deers} of different doublets.

Each of these contributions may be described by the product of a
coefficient function $\cal C$,
which is composed out of coupling constants and gauge boson propagators, and
 a kinematical function ${\cal G}(s,s_1,s_2)$.
The latter depends only on the three invariant masses and contains the
dynamics of the
corresponding hard scattering process.
The $\cal G$ functions, which are needed for a description of the {\tt
  CC11} process
are shown in table~\ref{tab3}.
The double resonating {\tt CC3} process is known for long to be
described by three
kinematical functions~\cite{muta}:
${\cal G}_{\tt CC3}^a, a={33,3f,ff}$.
The double resonating neutral current {\tt NC2} process, $e^+e^-
\rightarrow (ZZ) \rightarrow 4f$,
 proceeds via two diagrams of the {\tt crab} type and is characterized
 by only one kinematical
function ${\cal G}_{\tt NC2}$\footnote{
Since not only $Z$ pairs, but also $Z\gamma$ and $\gamma\gamma$
intermediate states exist one is faced in practice
with an {\tt NC8} process with the same topology like {\tt NC2}.
}~\cite{nc2}.
We also have to mention that the simplest final
state configuration of a complete neutral current $4f$ process has 24
Feynman diagrams and requires, besides ${\cal G}_{\tt NC2}$, only one
additional function ${\cal G}_{\tt NC24}$~\cite{nc24}.
It will be shown below that a complete description of the {\tt CC11}
process after the angular integrations is possible with adding
only one further function ${\cal G}_{\tt CC11}^{u,d}$.

\begin{table}[hbt]
\label{tab3}
\begin{center}
\begin{tabular}{|c||c||c|c||c|c|}
\hline 
&&&&&\\
process       &  $\sigma_{\tt CC3}$ & $\sigma^{3f}$ & $\sigma^{\nu f}$ &
$\sigma^{ff}$ & $\sigma^{f_1f_2}$
\\&&&&&
\\ \hline 
&&&&&\\
${\cal G}(s;s_1,s_2)$   
& ${\cal G}_{\tt CC3}^{33}$, ${\cal G}_{\tt CC3}^{3f}$, ${\cal G}_{\tt
  CC3}^{ff}$
& ${\cal G}_{\tt CC3}^{3f}$ & ${\cal G}_{\tt CC11}^{u,d}$;
${\cal G}_{\tt NC24}$
 & ${\cal G}_{\tt CC3}^{ff}$; ${\cal G}_{\tt NC2}$
& ${\cal G}_{\tt CC11}^{u,d}$; ${\cal G}_{\tt NC24}$
\\&&&&&
\\ \hline 
\end{tabular}
\caption[]
{\it
The complete set of kinematical functions describing the {\tt CC11} process.
}
\end{center}
\end{table}

In the next section,  notations and the formulae for the {\tt CC3}
process are introduced.
In section~3, the contributions from the background diagrams of figure~2 are
presented.
The photonic initial state and Coulomb corrections are added in section~4
and section~5 contains numerical results and a discussion.
Several appendices are devoted to technical details of the calculation.

\section{The {\tt CC3} process}
\ezero
\subsection{The total cross section}
The double resonating cross section is:
\ba
\sigma_{\tt CC3}(s;s_1,s_2)
&=&
\frac{\sqrt{\lambda}}{\pi s^2}
\biggl[
{\cal C}_{\tt CC3}^{s} {\cal G}_{\tt CC3}^{33}(s;s_1,s_2)
+
{\cal C}_{\tt CC3}^{st} {\cal G}_{\tt CC3}^{3f}(s;s_1,s_2)
+
{\cal C}_{\tt CC3}^{t} {\cal G}_{\tt CC3}^{ff}(s;s_1,s_2) \biggr]
{}.
\nl
\label{sigsstt}
\ea

The kinematical functions are known from~\cite{muta}:
\ba
{\cal G}_{\tt CC3}^{33}(s;s_1,s_2)
\label{gs4}
&=&
\frac{\lambda}{48} \left[ \lambda + 12\left(ss_1+s_1s_2+s_2s\right)\right],
\\
{\cal G}_{\tt CC3}^{{3f}}(s;s_1,s_2)
&=&
\frac{1}{48} \Biggl\{
(s-s_1-s_2)
\left[\lambda + 12s(s s_1 + s_1s_2 +s_2 s)\right]
\nl & &
-~24 \left[ ss_1+s_1s_2+s_2s\right] s_1s_2{\cal L}(s;s_1,s_2)\Biggr\},
\label{gst4}
\\
{\cal G}_{\tt CC3}^{ff}(s;s_1,s_2)
\label{gt4}
&=&\frac{1}{48}
\left[  \lambda + 12s(s_1+s_2) -48s_1s_2
+ 24 (s-s_1-s_2)s_1s_2 {\cal L}(s;s_1,s_2)\right].
\ea
In the superscript, the `3' indicates to the matrix element with a three
gauge boson vertex and `$f$' to the diagram with fermion propagator.
The function ${\cal G}_{\tt CC3}^{{3f}}$ has been slightly simplified
compared to~\cite{wwteup,muta}.
The logarithm in~(\ref{gst4}) and~(\ref{gt4}) arises from the
integration over the fermion propagator
in the $t$ channel (see appendix~\ref{psi}):
\ba
{\cal L}(s;s_1,s_2)
=
\frac{1}{\sqrt{\lambda}} \, \ln \frac{s-s_1-s_2+\sqrt{\lambda}}
                                     {s-s_1-s_2-\sqrt{\lambda}}.
\label{L}
\ea
Some properties of ${\lambda}$ and ${\cal L}$ are collected for the
convenience of the reader
in appendix~\ref{lamlog}.

For the coefficient functions, we choose generic definitions.
At first glance this seems to lead to artificially complicated
constructions.
At a later stage, however, the usefulness will become quite evident.
We begin with
\ba
{\cal C}_{\tt CC3}^{t}
&=&
\sum_{V_i,V_j,V_k,V_l=W}
{\cal C}_{422}^{ijkl|ee}(E,s;F_1,s_1;F_2,s_2).
\label{c422}
\ea
The coefficient ${\cal C}_{422}^{ijkl|ab}$ has been introduced for the neutral
current {\tt NC2} process in eq.~(11) of~\cite{nc24}; with a slight change
of notion for the current presentation, it becomes:
\ba
{\cal C}_{422}^{ijkl|ab} (F_2,s_2; F_1,s_1; E,s)
&=&
\frac{2}{(6\pi^2)^2} \, \,
\Re e
\frac{1}{D_{V_i}(s)D_{V_j}(s_1)D_{V_k}^*(s)D_{V_l}^*(s_1)}
\nl & & \times~
\biggl[ L(f_2^a,V_i) L(f_2^b,V_k) L(f_2^a,V_j) L(f_2^b,V_l)
\nl &&+~
R(f_2^a,V_i) R(f_2^b,V_k) R(f_2^a,V_j) R(f_2^b,V_l) \biggr]
N_c(F_2)
\nl & & \times~
\left[ L(F_1,V_j) L(F_1,V_l) +  R(F_1,V_j) R(F_1,V_l) \right] N_c(F_1)
\nl & &\times~
\left[ L(E,V_i) L(E,V_k) + R(E,V_i) R(E,V_k) \right] N_c(E).
\label{c422g}
\ea
The labels $E, F_1, F_2$ denote the corresponding members of weak
isodoublets.
For the charged current, all left- and right-handed fermion couplings
are equal:
\ba
L(e,W) \equiv L(E,W) = L(F_i,W) &=& \frac{g}{2\sqrt{2}},
\\
R(e,W) \equiv R(E,W) = R(F_i,W) &=& 0.
\label{gwff}
\ea
Below we will also need neutral current couplings where flavors will
cause differences for the couplings.

The boson propagator is:
\ba
D_V(s) = s-M_V^2+i\sqrt{s}\Gamma_V(s).
\label{gammaV}
\ea

With these definitions and using the relation
$G_{\mu}/\sqrt{2}=g^2/(8M_W^2)$, one easily verifies that
\ba
C_{\tt CC3}^t &=&
\frac{\left(G_{\mu} M_W^2 \right)^2}{s_1s_2}
\rho(s_1)
\rho(s_2),
\label{cc3t}
\ea
where
\ba
\rho(s_i)
&=&
\frac{1}{\pi}
\frac {\sqrt{s_i} \, \ww (s_i) }
      {|s_i - M_W^2 + i \sqrt{s_i} \, \ww (s_i) |^2} \times {\mr{BR}}(i),
\label{rho}
\ea
and
\bq
\ww (s_i)
=
 \sum_f \frac{G_{\mu}\, \wm ^2} {6 \pi \sqrt{2}}
 \sqrt{s_i}.
\label{gwoff}
\eq
The off shell width $\ww (s_i)$ is used throughout this paper.
It
contains a sum over all open
fermion decay channels $f$ at energy $\sqrt{s_i}$, and BR($i$) is the
corresponding branching ratio.
Equation~(\ref{cc3t})
makes the presence of the two Breit-Wigner factors explicit.
These are normalized such that
$\lim_{\Gamma_V\rightarrow 0} \rho(s_i) = \delta(s_i-M_W^2) \times BR(i)$.

The other two coefficient functions are:
\ba
{\cal C}_{\tt CC3}^{s}
&=&
\sum_{i,j=\gamma,Z} \frac{2}{\left(6\pi^2\right)^2} \Re e
\frac{1}{ \left|D_W(s_1)\right|^2\left|D_W(s_2)\right|^2
D_{V_i}(s) D_{V_j}^*(s)}
\nl &&\times~
g_3(V_i) g_3(V_j) L^2(F_1,W)  L^2(F_2,W) N_c(F_1) N_c(F_2)
\nl &&\times~
\left[L(e,V_i)L(e,V_j)  + R(e,V_i)R(e,V_j)\right] N_c(E),
\\
{\cal C}_{\tt CC3}^{st}
&=&
\sum_{j=\gamma,Z;k=W}
{\cal C}_{223}^{jk}(F_1,s_1;e,s;F_2,s_2),
\\
{\cal C}_{223}^{jk}(F_1,s_1;h,s_h;F,s_f)
&=&
\frac{2}{\left(6\pi^2\right)^2} \Re e
\frac{1}{ \left|D_W(s_1)\right|^2
D_W^*(s_2) D_{V_j}^*(s) D_{V_k}(s_f)}
\nl &&\times~
L^2(F_1,W) g_3(V_j)  L^2(F_2,W) L(h,V_k) N_c(F_1) N_c(F_2)
\nl &&\times~
\left[L(e,V_j)L(e,V_k)+R(e,V_j)R(e,V_k)\right] N_c(E),
\label{c223}
\ea
with
\ba
\begin{array}{rclrcl}
L(f,\gamma)&=&\frac{\displaystyle eQ_fg}{\displaystyle 2},
&
L(f,Z)&=&\frac{\displaystyle e}{\displaystyle
  4s_Wc_W}\left(2I_3^f-2Q_fs_W^2\right)  ,
\\
R(f,\gamma)&=& \frac{\displaystyle eQ_fg}{\displaystyle 2},
\hspace{.3cm}
&
\hspace{.3cm}
R(f,Z)&=&\frac{\displaystyle e}{\displaystyle
  4s_Wc_W}\left(-2Q_fs_W^2\right)  ,
\\
g_3(\gamma) &=& g s_W,
&
g_3(Z) &=& g c_W,
\label{couplr}
\end{array}
\ea
and
$e=gs_W=\sqrt{4\pi\alpha}$, $Q_e=-1, I_3^e=-\frac{1}{2}$.
\subsection{The angular distribution
\label{tad}
}
It is quite useful to know not only the invariant
mass distributions and the total cross section, but in addition also
the distribution in the production angle of one of the $W$ bosons:
\ba
\frac{d\sigma_{\tt CC3}}{d\cos\theta}
&=&
\frac{\sqrt{\lambda}}{\pi s^2}
\Biggl[
{\cal C}_{\tt CC3}^{s} {\cal G}_{\tt CC3}^{33}(s;s_1,s_2;\cos\theta)
\nl &&+~
{\cal C}_{\tt CC3}^{st} {\cal G}_{\tt CC3}^{3f}(s;s_1,s_2;\cos\theta)
+
{\cal C}_{\tt CC3}^{t} {\cal G}_{\tt CC3}^{ff}(s;s_1,s_2;\cos\theta)
\Biggr] ,
\label{sigstc}
\ea
with
\ba
\label{gcs}
{\cal G}_{\tt CC3}^{33}(s;s_1,s_2;\cos\theta)
&=&
\frac{1}{8}
 \left[\lambda C_1 + 12 s_1 s_2 C_2 \right],
\\     
\label{gcst}
{\cal G}_{\tt CC3}^{3f}(s;s_1,s_2;\cos\theta)
&=&\frac{1}{8}
\left[(s-s_1-s_2)C_1 - \frac{4s_1s_2[s(s_1+s_2)-C_2]}{t_{\nu}} \right],
\\         
\label{gct}
{\cal G}_{\tt CC3}^{ff}(s;s_1,s_2;\cos\theta)
&=&\frac{1}{8}
\left[ C_1 +  \frac{4s_1s_2C_2}{t_{\nu}^2} \right],
\ea
and
\ba
C_1&=&2s(s_1+s_2) + C_2,
\\
C_2&=&t_{\nu}(s-s_1-s_2-t_{\nu})-s_1s_2.
\ea
The scattering angle is contained in the denominator of the
$t$ channel propagator:
\ba
t_{\nu} = \frac{1}{2} \left( s-s_1-s_2 - \sqrt{\lambda} \cos\theta\right).
\ea

With the aid of appendix~\ref{psi}, it is trivial to integrate over
$\cos \theta$:
\ba
{\cal G}_{\tt CC3}^a(s,s_1,s_2)
= \frac{1}{2} \int_{-1}^1 d\cos\theta \,\,
{\cal G}_{\tt CC3}^a(s,s_1,s_2;\cos\theta), \hspace{.7cm}a=33,3f,ff.
\ea

The distribution~(\ref{sigstc}) is used for the description of $W$
pair production in {\tt PYTHIA}~\cite{pythia}.
It also may be used for the calculation of moments like
\ba
\int ds_1\, ds_2 \,d\cos\theta \,\, \cos^n \theta \,
\frac{d\sigma}{ds_1 ds_2 d\cos\theta}
\label{mom}
\ea
as a check on the accuracy of Monte Carlo integrations~\cite{wpairs}.
The angular distribution is more interesting in the context of
anomalous couplings of gauge bosons~\cite{anom}.
\section{Background contributions
\label{tbc}
}
\ezero
We now come to the contributions from the eight background diagrams of
figure~\ref{fig.2} and from their interferences with the double
resonating diagrams of figure~\ref{fig.1}.
For this purpose, we use a classification, which was introduced for
neutral current processes in~\cite{nc24}.
A single resonant diagram with a virtual fermion $f$ is called an
$f$-{\tt{deer}}.
Strictly speaking, the {\tt deers} are double resonating diagrams as
{\tt crabs} are:
besides the one $W$ resonance, they contain the $s$ channel $Z$ (or
$\gamma$) propagator.
One of the invariant masses, in which the diagram may become resonating,
is `eaten' by the fermion line.
In the {\tt crab} it is $s$, while in a  {\tt deer} either $s_1$ or $s_2$.
In this language, the {\tt crab} is an $e$-{\tt{deer}}.
In appendix~\ref{psi} it is made plausible that in fact this observation may
be used for a treatment of all the contributions on an equal footing.
\subsection{The {\tt crayfish}-{\tt{deer}} interferences $\sigma^{3f}$
\label{tcdi}
}
The interferences of the {\tt crayfish} diagram with the four types of
{\tt deers}
are similar to the {\tt crayfish}-{\tt{crab}} interference:
\ba
\sigma^{3f}
&=&
\frac{\sqrt{\lambda}}{\pi s^2}
\sum_{n=1,2}
\left[
{\cal C}_{\tt CC11}^{3u_n} + {\cal C}_{\tt CC11}^{3d_n} \right]
 {\cal G}_{\tt CC3}^{3f}(s_n;s,s_{3-n}).
\label{3f}
\ea

The coefficient function is
\ba
{\cal C}_{\tt CC11}^{3 f_n}
&=&
\sum_{V_j,V_k=\gamma,Z}
{\cal C}_{223}^{jk} (F_{3-n},s_{3-n}; f_n,s_n; e,s),
\ea
and ${\cal G}_{\tt CC3}^{3f}$ is defined in~(\ref{gst4}) and ${\cal
  C}_{223}^{jk}$
in~(\ref{c223}).
\subsection{The {\tt deer} interferences $\sigma^{ff}$
}
The contribution from the square of two {\tt deers}, which belong to one
doublet $F=(f^u,f^d)$ occurs twice:
\ba
\sigma^{ff}
&=&
\frac{\sqrt{\lambda}}{\pi s^2} \sum_{n=1,2}
\Biggl\{
\left[
{\cal C}_{\tt CC11}^{f_n^u f_n^u}
+ {\cal C}_{\tt CC11}^{f_n^df_n^d}
\right]
{\cal G}_{\tt CC3}^{ff} (s_n;s_{3-n},s)
+ {\cal C}_{\tt CC11}^{f_n^u f_n^d}
\, \,
{\cal G}_{\tt CC11}^{u,d} (s_n;s_{3-n},s)
\Biggr\} ,
\ea
with the coefficient function
\ba
{\cal C}_{\tt CC11}^{f_n^a f_n^b}
&=&
\sum_{V_i,V_k=\gamma,Z|V_j=V_l=W}
{\cal C}_{422}^{ijkl|ab} (F_n,s_n; F_m,s_m; E,s).
\label{fiafib}
\ea
The ${\cal C}_{422}^{ijkl|ab}$ is defined in~(\ref{c422g}).

The kinematical functions are different for the pure squares of diagrams,
where~(\ref{gt4})  is to be used, and for their interference.
The latter may be expressed by known functions.
Let us refer for a moment to the neutral current case.
There, all couplings are equal and one gets the following relation:
${\cal G}_{\tt NC2} = {\cal G}^{uu}+ {\cal G}^{dd}+ {\cal G}^{ud}$.
With ${\cal G}^{uu}= {\cal G}^{dd} = {\cal G}_{\tt CC3}^{ff}$, and
${\cal G}^{ud}
={\cal G}_{\tt CC11}^{ud}$, one gets the requested identity,
\ba
 {\cal G}_{\tt CC11}^{ud}(s;s_1,s_2)
=
{\cal G}_{\tt NC2}(s;s_1,s_2)
-
2  \, {\cal G}_{\tt CC3}^{ff} (s;s_1,s_2),
\label{g2udb}
\ea
with the neutral current
function ${\cal G}_{\tt NC2}$ from~\cite{nc2}; see appendix~\ref{gnc224}.
\subsection{The {\tt deer} interferences $\sigma^{f_1f_2}$
\label{tdi}
}
There are four interferences among {\tt deer} diagrams belonging to different
doublets:
\ba
\sigma^{f_1f_2}
&=&
\frac{\sqrt{\lambda}}{\pi s^2}
\Biggl\{
\left[ {\cal C}^{f_1^u f_2^u} + {\cal C}^{f_1^d f_2^d} \right]
{\cal G}_{\tt CC11}^{uu,dd} (s;s_1,s_2)
+~
\left[ {\cal C}^{f_1^u f_2^d} + {\cal C}^{f_1^d f_2^u}  \right]
 {\cal G}_{\tt CC11}^{u,d} (s;s_1,s_2) \Biggr\} .
\label{sigF1}
\ea
The coefficient function may be traced back to that introduced in~\cite{nc24}:
\ba {\cal C}^{f_1^a f_2^b} &=& \sum_{V_i,V_k=\gamma,Z|V_j,V_l=W} {\cal
  C}_{233}^{ijkl} (E,s; f_1^a,s_1; f_2^b,s_2).
\label{fiafif}
\ea
The function ${\cal C}_{233}^{ijkl}$ is defined as follows:
{\small
\ba
\label{c233g}
{\cal C}_{233}^{ijkl}(E,s;F_1,s_1;F_2,s_2)
&=& \frac{2}{(6\pi^2)^2}
\;
\Re e
\frac{1}{D_{V_i}(s) D_{V_k}^*(s) D_{V_l}(s_1)D_{V_j}^*(s_2)}
\nl &  
\times&
\left[ L(E,V_i)L(E,V_k)+R(E,V_i)R(E,V_k)\right] N_c(E)
\nl &  
\times&
\left[ L(F_1,V_l) L(F_1,V_j) L(F_1,V_k) - R(F_1,V_l) R(F_1,V_j) R(F_1,V_k)
\right] N_c(F_1)
\nl &  
\times&
\left[ L(F_2,V_l) L(F_2,V_j) L(F_2,V_i) - R(F_2,V_l) R(F_2,V_j) R(F_2,V_i)
\right] N_c(F_2).
\nl
\label{cijkl}
\ea
} 
Wherever a neutral gauge boson couples, the corresponding fermion $e$
or $f_2$ etc. will replace the doublets $E$, $F_2$ in the arguments of
the couplings $L,R$.

For the kinematical functions, again a relation to a neutral current
function may be established.
The function ${\cal G}_{\tt NC24}$, explicitly given in
appendix~\ref{gnc224}, describes a
 sum of interferences analogue to those considered here.
Since the couplings are equal in that case, we have:
${\cal G}_{\tt NC24} = {\cal G}^{u_1u_2} + {\cal G}^{u_1d_2} + {\cal
  G}^{d_1u_2} +
{\cal G}^{d_1d_2}$.
Furthermore, using the symmetry relations
${\cal G}^{u_1u_2} =  {\cal G}^{d_1d_2} \equiv {\cal G}_{\tt CC11}^{uu,dd}$
and
${\cal G}^{u_1d_2} =  {\cal G}^{d_1u_2} \equiv {\cal G}_{\tt CC11}^{u,d}$,
we may write
${\cal G}_{\tt NC24} = 2 {\cal G}_{\tt CC11}^{uu,dd} + 2 {\cal G}_{\tt
  CC11}^{u,d}$.
This relation allows to determine
${\cal G}_{\tt CC11}^{uu,dd}$
once
${\cal G}_{\tt CC11}^{u,d}$
is known:
\ba
{\cal G}_{\tt CC11}^{uu,dd}
=
\frac{1}{2} {\cal G}_{\tt NC24}
-
{\cal G}_{\tt CC11}^{u,d}.
\ea

By an explicit calculation, we obtained:
\ba
\label{gnuu}
&&
{\cal G}_{\tt CC11}^{u,d} (s;s_1;s_2)
=
-120
s^4 \frac{s_1^3 s_2^3}{\lambda^3}
   {\cal L}(s_2;s,s_1) {\cal L}(s_1;s_2,s) \nl
 &&\hspace{1.2cm}
-s
\Biggl[1+\frac{s(s-\sigma)}{\lambda}
     +20s^2\frac{s_1s_2}{\lambda^2}
   -30s^3s_1s_2\frac{s-3\sigma}{\lambda^3}\Biggr]
 \Biggl[s_1^2 {\cal L}(s_2;s,s_1)+s_2^2 {\cal L}(s_1;s_2,s)\Biggr]\nl
 &&\hspace{1.25cm}
-s(s_1-s_2)
\Biggl[\frac{s-\sigma}{\lambda}
   +10s\frac{s_1s_2}{\lambda^2}
 -30s^2s_1s_2\frac{s+\sigma}{\lambda^3}\Biggr]
 \Biggl[s_1^2 {\cal L}(s_2;s,s_1)-s_2^2 {\cal L}(s_1;s_2,s)\Biggr]\nl
 &&\hspace{1.2cm}
-\frac{1}{12}
\Biggl\{(s^2-\sigma^2)
 \Biggl[1+12\frac{s\sigma}{\lambda}-60s^2\frac{s_1s_2}{\lambda^2}\Biggr] \nl
 &&\hspace{2.67cm} -8s_1s_2\Biggl[1-\frac{s(4s+5\sigma)}{\lambda}
  +15s^2\frac{s_1s_2}{\lambda^2}\Biggl(1-6\frac{s(s-\sigma)}{\lambda}
 \Biggr)\Biggr]\Biggr\},
\ea
where $\sigma = s_1 + s_2$.

For numerical applications, it is helpful to know the limit
$\lambda \rightarrow 0$:
\ba
\lim_{\lambda \rightarrow 0}
{\cal G}_{\tt CC11}^{u,d} (s;s_1;s_2)
&=&
-\frac{1}{64}\left(s^2-\Delta^2\right)\left(9-\frac{\Delta^2}{s^2}\right)
+ \frac{
{{\lambda}}
}{160} \left(1+9\frac{\Delta^2}{s^2}\right) + {\cal O}
(
{{\lambda}}
^2),
\label{limgud}
\ea
where $\Delta = s_1-s_2$.
\subsection{The {\tt crab}-{\tt{deer}} interferences $\sigma^{\nu f}$
}
There are four {\tt crab}-{\tt{deer}}
interferences:
\ba
\sigma^{\nu f}
&=&
\frac{\sqrt{\lambda}}{\pi s^2}
\sum_{n=1,2}
\left[
{\cal C}_{\tt CC11}^{\nu_e u_n}
\, \, {\cal G}_{\tt CC11}^{u,d} (s_{3-n};s,s_n)
+
{\cal C}_{\tt CC11}^{\nu_e d_n}
\, \, {\cal G}_{\tt CC11}^{uu,dd} (s_{3-n};s,s_n)
\right].
\label{nuf}
\ea

The coefficient functions are:
\ba
{\cal C}_{\tt CC11}^{\nu_ef_n}
&=&
\sum_{V_i=\gamma,Z|V_j,V_k,V_l=W} {\cal
  C}_{233}^{ljki}(F_{3-n},s_{3-n};E,s,f_n,s_n),
\label{cef}
\ea
and the ${\cal C}_{233}^{ljki}$ is defined in~(\ref{c233g}).

The kinematical functions are introduced in section~\ref{tdi}.
One should be aware of a factor $\frac{1}{2}$
resulting in the limit of neutral couplings.
Differing from the neutral current case, there is no charged current
crossed {\tt crab} diagram, which would give the same contribution as
the {\tt crab} diagram.
%
\section{QED corrections
\label{qed}}
\ezero
A complete treatment of QED corrections for four-fermion production is
part of electroweak corrections and a quite ambitious
task~\cite{vanO}.
For the time being, one may try to restrict to initial state radiation
(ISR) and, near the $W$ pair production threshold, corrections from
the Coulomb singularity.
Both corrections influence the cross section significantly.
%
\subsection{Flux function approach
\label{ffa}
}
%

The cross section with QED corrections becomes in the flux function approach:
\ba
\frac{d\sigma_{QED}(s)}{ds_1ds_2}
&=&
\int
\limits_{\left(\sqrt{s_1}+\sqrt{s_2}\right)^2}^s \frac{ds'}{s}
\Biggl\{
G(s'/s) \Bigl[ \sigma_{\tt CC11}(s',s_1,s_2)
+~C(s') \sigma_{\tt CC3}(s',s_1,s_2) \Bigr]
\nl &&
+~ \sigma_{QED}^{non-univ}(s',s_1,s_2) \Biggr\},
\label{sigff}
\ea
with
\ba
G(s'/s)
&=&
\beta_e \left(1-\frac{s'}{s}\right)^{\beta_e-1} (1+S) + H(s',s),
\ea
and
\ba
\label{beta.e}
\beta_e&=&\frac{2\alpha}{\pi}\left(L_e-1\right),
\\
\label{le}
L_e &=& \ln\frac{s}{m_e^2}.
\ea
The virtual and soft corrections $S$ and hard corrections $H$
are~\cite{berends}:
\ba
S
&=&
\frac{3}{4} \beta_e + \frac{\alpha}{\pi}
\left(\frac{\pi^2}{3}-\frac{1}{2}\right) \times {\tt IZERO}
+ S^{(2)},
\\
S^{(2)}&=&
\left(\frac{\alpha}{\pi}\right)^2 \left[ s_2 L_e^2 + s_1 L_e + s_0 \right]
+ S_{pairs}^{(2)},
\\
H(s',s)&=&-\frac{1}{2}\left(1+\frac{s'}{s}\right) \beta_e + H^{(2)},
\\
H^{(2)}&=&
\left(\frac{\alpha}{\pi}\right)^2 \left[ h_2 L_e^2 + h_1 L_e + h_0 \right]
+ H_{pairs}^{(2)},
\ea
where the corrections $S_{pairs}^{(2)}$ and $H_{pairs}^{(2)}$ due to
fermion pair emission from the initial
state have been determined in~\cite{kuehn}.

In the leading logarithmic approximation (LLA), the ISR is well defined.
The complete ${\cal O}(\alpha)$ corrections include additional
contributions in case of $t$ channel exchanges.
They have been defined and determined for the double resonating
{\tt CC3} (\cite{wwqed}) and {\tt NC2} (\cite{zzqed}) processes.
For a complete treatment of the {\tt CC11} process, this is not
sufficient and has to be accomplished by a proper treatment of the
{\tt crayfish}-{\tt{deer}} interferences;
these are, however, small and the non-leading QED corrections to them the more.

The Coulomb corrections apply to the double resonating diagrams~\cite{coulomb}:
\begin{eqnarray}
C(s)
&=&
\frac{\alpha(2 M_W)}{\beta} \, \Im m \ln
\left[ \frac
{\beta_W + \delta - \beta}
{\beta_W + \delta + \beta} \right],
\\
\beta &=& \frac{1}{s} \sqrt{\lambda},
\\
\beta_W &=& \frac{1}{s} \sqrt{\lambda\left(s,m_W^2,m_W^2\right)},
\\
m_W^2 &=& M_W^2-iM_W\Gamma_W,
\\
\delta &=& \frac{1}{s}\left|s_1^2-s_2^2\right|.
\end{eqnarray}

Besides the cross section, there is also experimental interest to know the
radiative energy loss due to ISR.
This quantity cannot be calculated in the flux function approach.
However, the calculation of a similar quantity, namely the
invariant mass loss (of final state
fermions) is possible:
\begin{eqnarray}
\langle m_{\gamma} \rangle
&=&
\frac{1}{\sigma}
\int ds_1 ds_2 \,
\int \frac{ds'}{s} \frac{\sqrt{s}}{2}\left(1-\frac{s'}{s}\right)
\frac{d\sigma}{ds_1ds_2ds'},
\end{eqnarray}
where $d\sigma/ds_1ds_2ds'$ is the contents of the curly brackets
in~(\ref{sigff}).

%
\subsection{Structure function approach
\label{sfa}
}
%
Alternatively, the
structure function approach may be realized with the following formula:
\ba
\frac{d\sigma_{QED}(s)}{ds_1ds_2}
&=&
\int
 \limits_{x_1^{\min}}^1 dx_1
\int
 \limits_{x_2^{\min}}^1 dx_2
\, D(x_1,s) D(x_2,s)
\Biggl\{
\sigma_{\tt CC11}(x_1x_2s,s_1,s_2)
\nl && +~
C(x_1x_2s)\sigma_{\tt CC3}(x_1x_2s,s_1,s_2)\Biggr\}
{}.
\label{sigsf}
\ea
The lower integration boundaries are:
\ba
x_1^{\min} &=& \frac{(\sqrt{s_1}+\sqrt{s_2})^2}{s},
\\
x_2^{\min} &=& \frac{(\sqrt{s_1}+\sqrt{s_2})^2}{x_1s}.
\ea
To a very good accuracy, the structure function may be written as
follows~\cite{kufa}:
\ba
D(x,s)
&=&
(1-x)^{{\beta_e}/{2}-1} \, \frac{\beta_e}{2} (1+S) + H(x,s),
\label{dxs}
\ea
and
\ba
S &=&
\frac{e^{({3}/{4}-\gamma_E){\beta_e}/{2}}}{\Gamma\left(1+\frac{\beta_e}{2}
\right)}
- 1,
\\
H(x,s) &=& -\frac{1}{2} (1+x) \frac{\eta_e}{2}
\nl && +~ \frac{1}{8} \left[ -4 (1+x)\ln(1-x) + 3 (1+x)\ln x - 4
\frac{\ln x }{1-x} -5 -x \right]  \left(\frac{\eta_e}{2}\right)^2,
\\
\eta_e &=& \frac{2\alpha}{\pi} \left( L_e - {\tt IZETTA}\right).
\ea

Within a given order of LLA, the structure function and the flux
function approaches are related:
\begin{eqnarray}
G(s'/s) &=&
\int dx_1dx_2 \, D(x_1,s) D(x_2,s) \,  \delta\left(
x_1x_2-\frac{s'}{s}\right).
\end{eqnarray}

The radiative energy loss is determined as follows:
\begin{eqnarray}
\langle E_{\gamma} \rangle
&=&
\frac{1}{\sigma}
\int ds_1 ds_2
\int dx_1dx_2 \, D(x_1,s) D(x_2,s) \frac{\sqrt{s}}{2}[(1-x_1)+(1-x_2)]
\frac{d\sigma}{ds_1 ds_2dx_1dx_2},
\nll
\end{eqnarray}
and the invariant mass loss (of final state fermions) is:
\begin{eqnarray}
\langle m_{\gamma} \rangle
&=&
\frac{1}{\sigma} \int ds_1 ds_2
\int dx_1dx_2 \, D(x_1,s) D(x_2,s) \frac{\sqrt{s}}{2}(1-x_1x_2)
\frac{d\sigma}{ds_1 ds_2dx_1dx_2},
\end{eqnarray}
where $d\sigma/ds_1 ds_2dx_1dx_2$ is the content of the curly
bracket in~(\ref{sigsf}).

In table~\ref{fsig}, we have collected the flags, which are used in
{\tt GENTLE} in order to define the details of the calculations.

\begin{table}[thbp]
\begin{center}
{\begin{tabular}[]{|c|c|c|c|c|}
\hline 
\hspace{.5cm} Flag \hspace{.5cm} & FF & SF & &
\\
\hline 
\hline
{\tt ICONVL}    &   0    & 1    &   &
\\ \hline 
{\tt ICOLMB}    & 0 -- 5 & --   & 3 & 0
\\ \hline 
{\tt IQEDHS}    & 0 -- 4 & --   & 3 & 2
\\ \hline 
{\tt IZERO}     & 0, 1   & --   & 1 & 0
\\ \hline 
{\tt IZETTA}    & --     & 0, 1 & 1 & 1
\\ \hline 
{\tt ITVIRT}    & 0, 1   & --   & 1 & 0
\\ 
{\tt ITBREM}    & 0, 1   & --   & 1 & 0
\\ \hline 
\end{tabular}
} 
\caption{
\it
Some of the flag settings in {\tt GENTLE} for the flux function (FF)
and the structure function (SF) approaches.
The third column of flags contains recommended flag values for best
estimates and the last one those chosen for the comparisons in
table~6.
\label{fsig}
}
\end{center}
\end{table}

%
\section{Numerical results and discussion\label{numeric}}
\ezero
%
%
The numerical results are obtained with the Fortran program {\tt
  GENTLE}~\cite{gentle}.
The input quantities are chosen as follows:
\ba
G_{\mu}  &=& 1.16639\times10^{-5}~{\rm GeV}^2,
\nl
\alpha (2 M_W) &=& (128.07)^{-1},
\nl
M_Z    &=&  91.1888~{\rm GeV},
\nl
M_W   &=& 80.23~{\rm GeV}.
\ea
Further, for the QED corrections,
\ba
\alpha &=& (137.03599)^{-1} .
\ea
The $s$ dependent widths are $\Gamma_V(s) = (\sqrt{s}/M_V) \Gamma_V$.
For the $Z$ width we use the value
\ba
\Gamma_Z  = 2.4974~{\rm GeV},
\ea
thus circumventing a dependence of all the numbers on the top quark
mass from a prediction of $\Gamma_Z$ in the Standard
Model~\cite{zdecay}.
For the $W$ width the radiative corrections are small~\cite{wdecay}
and the following approximation is used:
\ba
\Gamma_W  = 9 \times \frac{G_{\mu} M_W^3}{6\pi\sqrt{2}}.
\ea
Finally, we need the weak mixing angle $\sin^2\theta_W$.
Again, we deviate from the pure Standard Model relation and use
instead an effective weak mixing angle:
\ba
s_W^2 = \frac{\pi \alpha(2 M_W)}{\sqrt{2}M_W^2G_{\mu}}.
\ea

\bigskip

In table~\ref{tcsia}, cross section predictions are shown for the
different channels of table~\ref{tbot} over a large range of
$\sqrt{s}$.
They are normalized by taking off the branching factors.
The differences between the channels are seen to be minor above
$\sqrt{s} \sim 160 $ GeV, where the double resonating {\tt CC3}
diagram are dominating.
Nevertheless, the predictions for the {\tt CC3} process and any of the
{\tt CC11} processes start to deviate substantially above $\sqrt{s} =
200$ GeV.

\begin{table}[hbt]
\begin{center}
\begin{tabular}{|l|l|c|}
\hline 
$d_1 {\bar u}_1$
&
\raisebox{0.pt}[2.5ex][0.0ex]{${\bar d}_2 u_2$}
&
\mbox{BR(1)}$\times$\mbox{BR(2)}
\\ \hline 
\hline
$l\nu$    & $l\nu$    & 1/81
\\ \hline 
$l\nu$    & $q^d q^u$ & 1/27
\\ \hline 
$q^d q^u$ & $l\nu$    & 1/27
\\ \hline 
$q^d q^u$ & $q^d q^u$ & 1/9
\\ \hline 
all       & all       & 1
\\ \hline 
\end{tabular}
\caption[]
{\it
The branching ratios of the different types of final states of the
{\tt CC11} process.
\label{tbot}
}
\end{center}
\end{table}

\begin{table}[hbt]
\begin{center}
\begin{tabular}{|c|r@{.}l|r@{.}l|r@{.}l|r@{.}l|r@{.}l|}
\hline 
$\sqrt{s}$
          &\multicolumn{2}{c|}{\tt CC3        }
          &\multicolumn{2}{c|}{$L \nu l \nu'$ }
          &\multicolumn{2}{c|}{$l \nu qq'$   }
          &\multicolumn{2}{c|}{$QQ'qq'$      }
          &\multicolumn{2}{c|}{\tt CC11}
\\ \hline 
\hline
30
&  \multicolumn{2}{c|}{1.4519$\times 10^{-7}$}  &  5&9295$\times
10^{-6}$   &  7&2478$\times
10^{-6}$   &   5&0897$\times 10^{-6}$   & 6&1422$\times 10^{-6}$
\\ \hline 
60
&  \multicolumn{2}{c|}{1.9358$\times 10^{-5}$} &  4&0025$\times
10^{-5}$ &  4&6879$\times
10^{-5}$ &  3&5441$\times 10^{-5}$ &4&1034$\times 10^{-5}$
\\ \hline 
91.189   &  0&11225  &  0&021329  &  0&024551  &  0&018975 & 0&021715
\\ \hline 
176  &16&225 &16&242 &16&243 &16&243& 16&243
\\ \hline 
200  &18&578 &18&586 &18&588 &18&588&  18&588
\\ \hline 
500  &        7&3731    &7&3301 &7&3318 &7&3334 & 7&3323
\\ \hline 
1000         &2&9888   &2&9342 &  2&9344&2&9348 & 2&9347
\\ \hline 
2000         &1&5702 &1&5020 &1&5018 &1&5016 & 1&5017
\\ \hline 
\end{tabular}
\caption[]
{\it
Total cross sections in nbarn for the different Born 4$f$ production
channels as functions of $\sqrt{s}$ (in GeV). For this comparison, the
branching ratios of table~4 are {\em not} taken into account in the
single mode channels.
\label{tcsia}
}
\end{center}
\end{table}


\newpage

Tables~\ref{rfdp} and~\ref{rfdp2} compare for LEP~2 and
table~\ref{rfdp4} for linear collider (LC) energies the numerical
predictions  of {\tt   GENTLE} with
those of the Monte Carlo programs {\tt WPHACT}~\cite{wphact}, {\tt
  WWGENPV}~\cite{wwgenpv} and {\tt WTO}~\cite{wto}; the latter
performs the numerical integrations with a deterministic approach.
{\tt GENTLE} may be used with both the flux function (FF) and the
structure function (SF) approaches, while the other programs are
based on the structure function approach.
Shown are $\sigma_{Born}$, $\sigma_{QED}$,  $\langle E_{\gamma} \rangle$, and
$\langle m_{\gamma} \rangle$
as being introduced in section~\ref{qed}.
For the observables with QED corrections, the first two lines at each
energy show that both treatments of QED corrections agree well.
Further, the predictions of the different programs are in perfect
agreement with each other.
Having in mind that some numbers arise from ninefold numerical
integrations, this is a nontrivial result proving the high level of
technical precision reached by
now with quite different techniques.
The dependence of cross sections on the channels has with QED
corrections the same tendencies as in the Born case.
As an example for this we mention that $\langle m_{\gamma} \rangle$ in
table~\ref{rfdp2} shows, within the digits shown, no channel
dependence at all.

\begin{table}[thbp]
\begin{center}
\begin{tabular}{|c|c|r@{.}l|r@{.}l|r@{.}l|}
\hline 
 & \hspace{.5cm}  $\sqrt{s}$ \hspace{.5cm}
&\multicolumn{2}{c|}{\hspace{.5cm}   176 \hspace{.5cm} }
&\multicolumn{2}{c|}{\hspace{.5cm}   190 \hspace{.5cm} }
&\multicolumn{2}{c|}{\hspace{.5cm}   205 \hspace{.5cm} }
\\
\hline 
\hline
$\sigma_{Born}$        & {\tt GENTLE}    &1&8048 &2&0405&2&0573
\\                     & {\tt WPHACT}    &1&8048(1)
&2&0405(1)&2&0574(1)
\\                     & {\tt WWGENPV}   &1&8048(1)
&2&0405(1)&2&0574(1)
\\ \hline 
\hline
$\sigma_{QED}$         & {\tt GENTLE,FF} &1&5067&1&8148&1&9022
\\                     & {\tt GENTLE,SF} &1&5046&1&8124&1&8998
\\                     & {\tt WPHACT}
&1&5046(1)&1&8124(1)&1&8999(1)
\\                     & {\tt WTO}
&1&5046(1)&1&8124(3)&1&8996(4)
\\                     & {\tt WWGENPV}
&1&5045(1)&1&8124(1)&1&9000(1)
\\ \hline 
\end{tabular}
\caption[]{
\it
Results from different programs for the total {\tt CC11}
cross sections  $e^+e^- \rightarrow 4f$ and
$e^+e^- \rightarrow 4f + \gamma $  (both in nbarn) at LEP~2 energies
for the $QQ' \, qq'$ channel.
The MC error estimates are indicated.
\label{rfdp}
}
\end{center}
\end{table}

\begin{table}[thbp]
\begin{center}
 \begin{tabular}  {|c|c|r@{.}l|r@{.}l|r@{.}l|}
\hline 
 & \hspace{.5cm}  $\sqrt{s}$ \hspace{.5cm}
&\multicolumn{2}{c|}{\hspace{.5cm}   176 \hspace{.5cm} }
&\multicolumn{2}{c|}{\hspace{.5cm}   190 \hspace{.5cm} }
&\multicolumn{2}{c|}{\hspace{.5cm}   205 \hspace{.5cm} }
\\
\hline 
\hline
$\langle E_{\gamma} \rangle$ & {\tt GENTLE,SF} & 1&1151& 2&1400&
3&2025
\\                           & {\tt WPHACT} &  1&1149(2) & 2&1402(4) &
3&2030(6)
\\                           & {\tt WTO}    & 1&1149(3)& 2&1404(5)&
3&2023(8)
\\                           & {\tt WWGENPV}& 1&1147(4)& 2&1400(5)&
3&2028(10)
\\ \hline 
\hline
$  \langle m_{\gamma} \rangle$  & {\tt GENTLE,FF}    & 1&1129&2&1310 &3&1816
\\                              & {\tt WWGENPV}& 1&1124(3) & 2&1316(5)
& 3&1861(11)
\\ \hline 
\end{tabular}
\caption[]{
\it
Results from different programs for
the average energy loss $\langle E_{\gamma} \rangle$ and the
invariant mass loss $\langle m_{\gamma} \rangle$ (both in GeV) due to
QED radiative corrections at LEP~2 energies
in the $l \nu \, qq'$ channel.
\label{rfdp2}
}
\end{center}
\end{table}

\begin{table}[bhtp]
\begin{center}
\begin{tabular}{|c|c|r@{.}l|r@{.}l|r@{.}l|}
\hline 
 & \hspace{.5cm}  $\sqrt{s}$ \hspace{.5cm}
&\multicolumn{2}{c|}{\hspace{.5cm}   500 \hspace{.5cm} }
&\multicolumn{2}{c|}{\hspace{.5cm}   1000 \hspace{.5cm} }
&\multicolumn{2}{c|}{\hspace{.5cm}   2000 \hspace{.5cm} }
\\
\hline 
\hline
$\sigma_{Born}$        & {\tt GENTLE}
&  0&81482   & 0&32609    & 0&16684
\\                     & {\tt WWGENPV}
&  0&81480(6)& 0&32602(6) & 0&16682(7)
\\ \hline 
\hline
$\sigma_{QED}$
&{\tt GENTLE,SF}
& 0&86950     &0&36514    &   0&18247
\\                     & {\tt WPHACT}
& 0&86956(9)  &0&36515(5) &   0&18250(4)
\\                     & {\tt WTO}
& 0&86960(25) & \multicolumn{2}{c|}{} & \multicolumn{2}{c|}{}
\\                     & {\tt WWGENPV}
& 0&86956(14) & 0&36530(35) & 0&18247(13)
\\ \hline 
\end{tabular}
\caption[]{
\it
Results from different programs for the total {\tt CC11}
cross sections  $e^+e^- \rightarrow 4f$ and
$e^+e^- \rightarrow 4f + \gamma $  (both in nbarn)
at Linear Collider energies for the $QQ' \, qq'$ channel.
The MC error estimates are indicated.
\label{rfdp4}
}
\end{center}
\end{table}

\newpage

\begin{figure}[hb]
\begin{center}
\vspace*{2.3cm}
\mbox{
\epsfysize=20.0cm
\epsffile{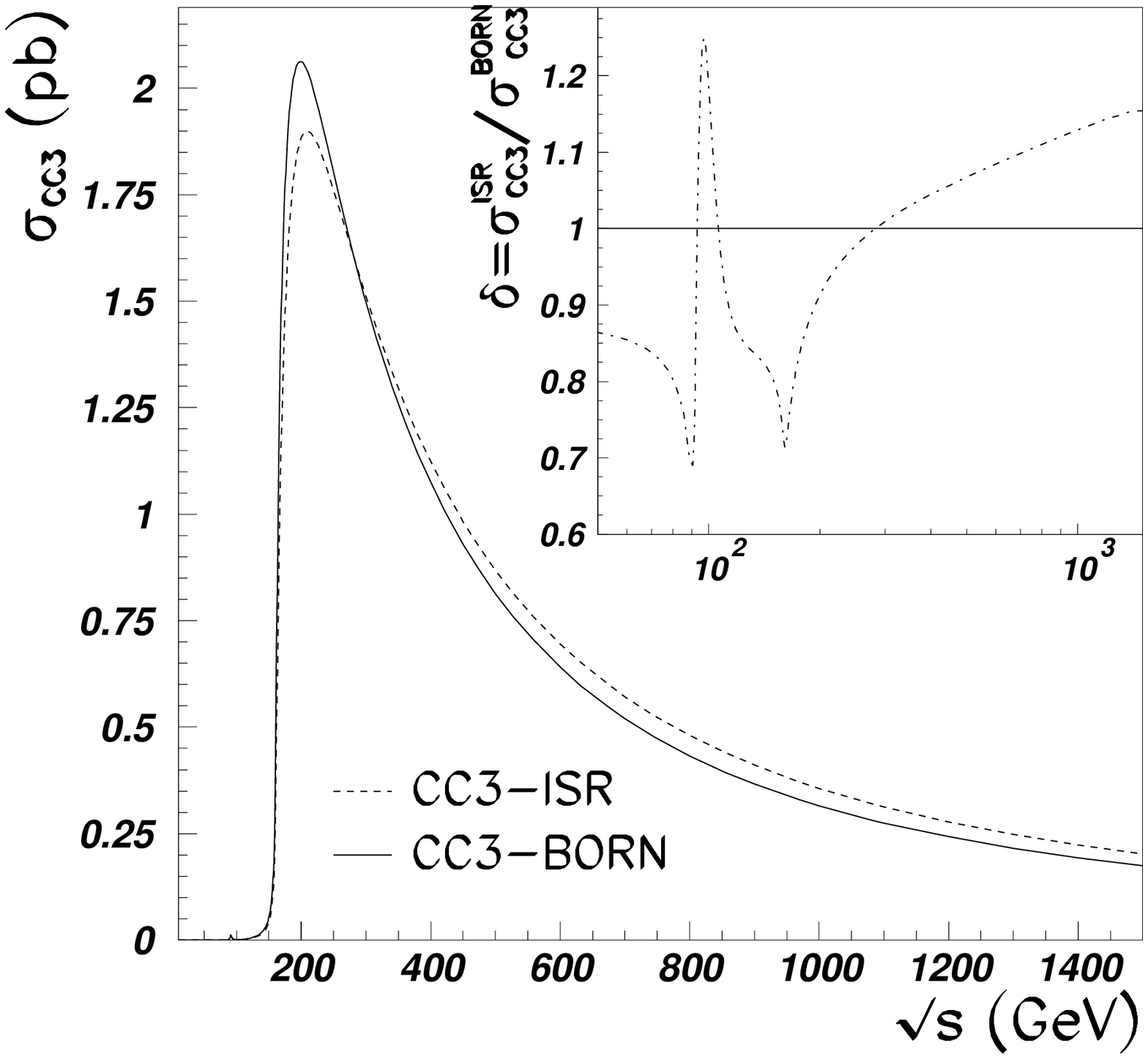}
}
\end{center}
\vspace*{-2.2cm}
\caption
{\it
The {\tt CC3} process without and with QED corrections.
}
\label{fcc3}
\end{figure}

Figure~\ref{fcc3} shows, over a large energy range, the {\tt CC3}
process without and with QED corrections.
The latter smear out the $W$ pair
 production peak and produce a radiative tail at higher energies,
 crossing over the Born curve at $\sqrt{s} \sim 240$  GeV.
The $Z$ resonance in the {\tt crayfish} diagram leads to an
interference pattern around $\sqrt{s} \sim 91$ GeV, with a set-on of
the $Z$ radiative tail until the $W$ pair production starts to
over-compensate it and to completely  dominate the process.

\begin{figure}[hb]
\begin{center}
\vspace{2.3cm}
\mbox{
\epsfysize=20.0cm
\epsffile{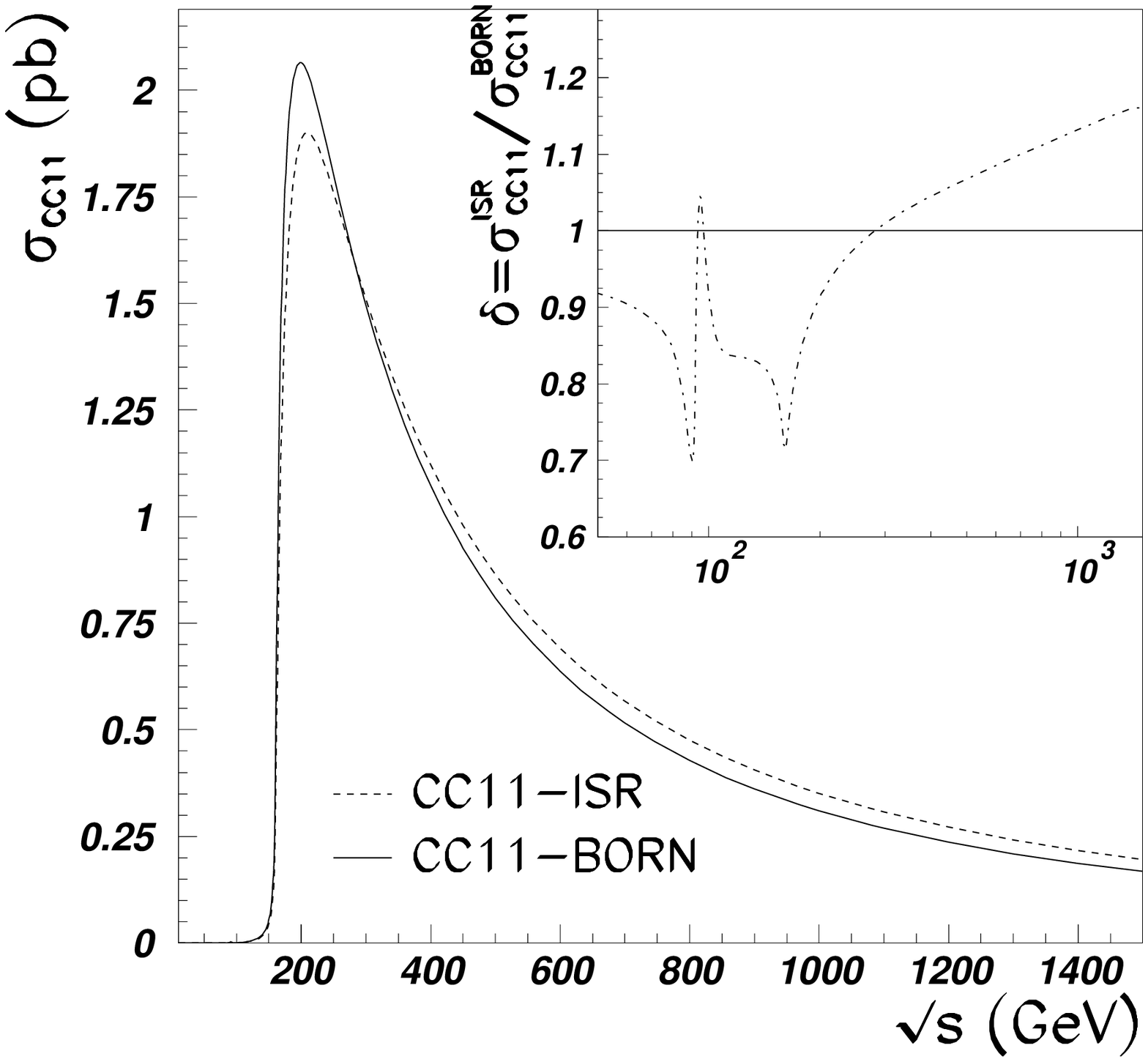}
}
\end{center}
\vspace{-2.2cm}
\caption
{\it
The {\tt CC11} process without  and with QED corrections.
}
\label{fcc11}
\end{figure}

The same, but now for the  {\tt CC11} process, is shown in figure~\ref{fcc11}.
As to be expected, the differences to the foregoing figure are not pronounced,
although visible.
Both the $Z$ and $W$ pair peaks (seen in the set-in as minima in
$\sigma_{QED}/\sigma_{Born}$) are nearly untouched, while the tail of
the $Z$ between the two minima gets suppressed due to the background.

\begin{figure}[hb]
\begin{center}
\vspace{2.3cm}
\mbox{
\epsfysize=20.0cm
\epsffile{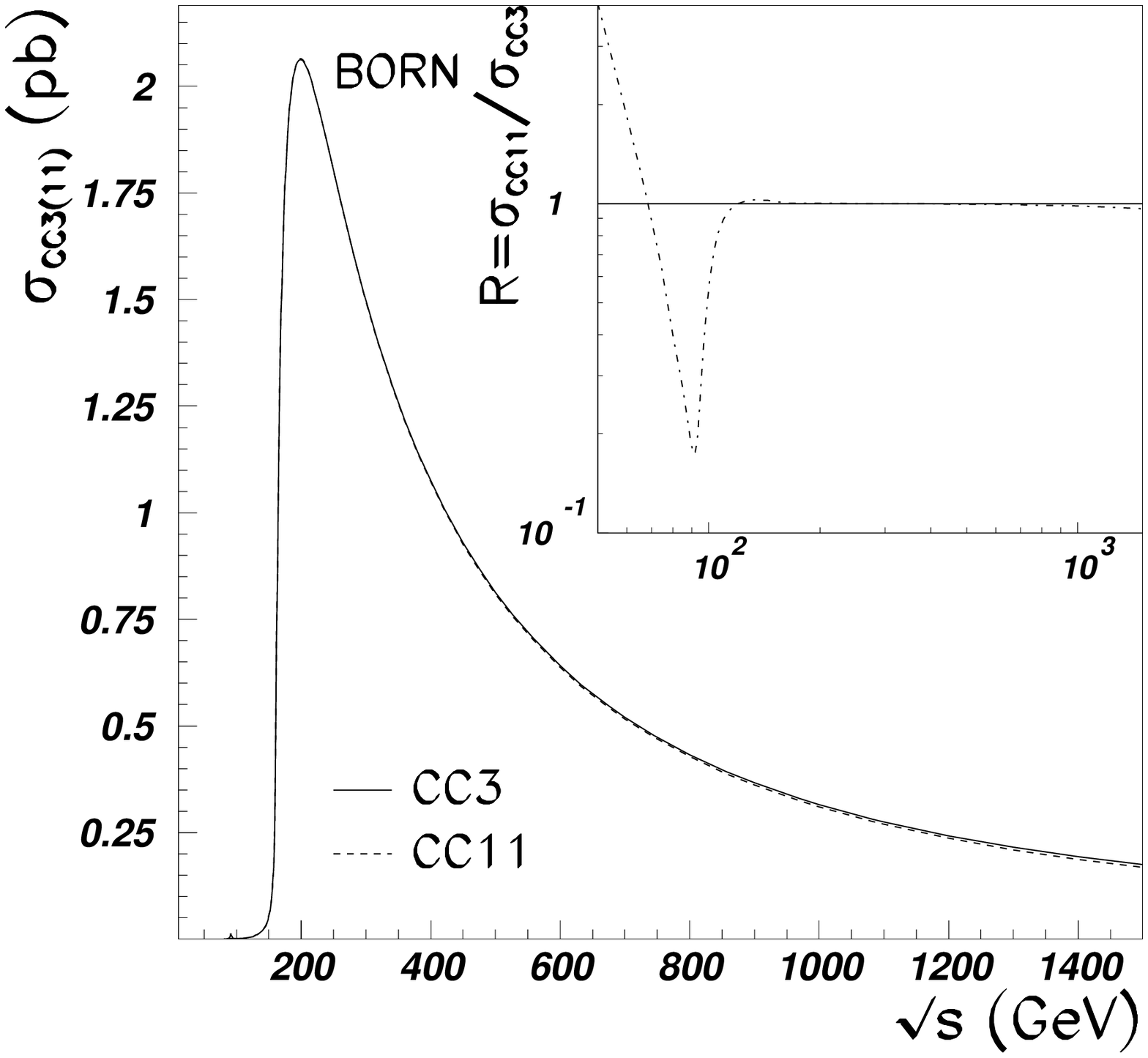}
}
\end{center}
\vspace{-2.2cm}
\caption
{\it
The ratio of {\tt CC11} and {\tt CC3} processes without QED corrections.
}
\label{frat311}
\end{figure}

\begin{figure}[hb]
\begin{center}
\vspace{2.3cm}
\mbox{
\epsfysize=20.0cm
\epsffile{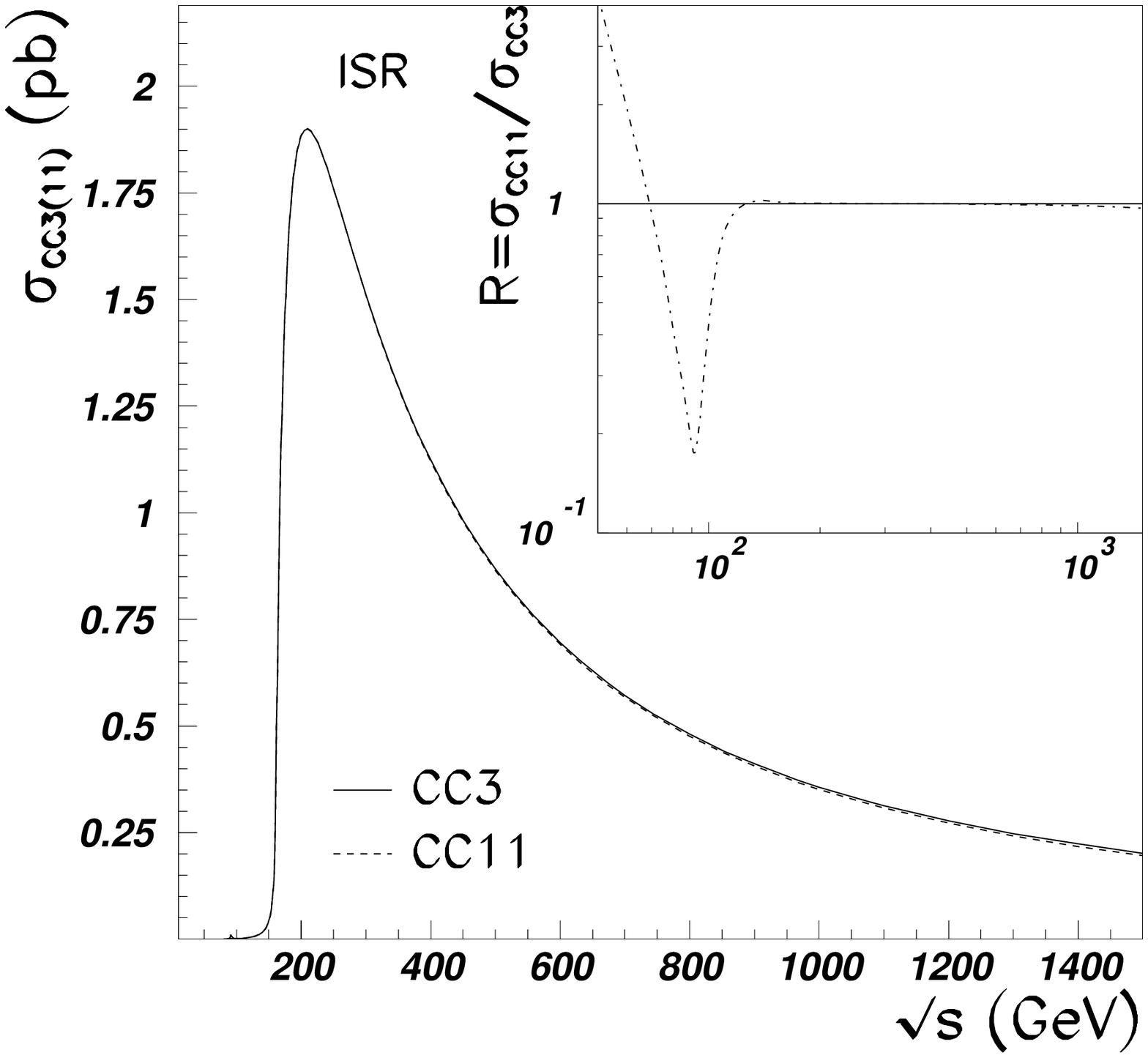}
}
\end{center}
\vspace{-2.2cm}
\caption
{\it
The ratio of {\tt CC11} and {\tt CC3} processes with QED corrections.
}
\label{frat311i}
\end{figure}

Direct comparisons of the {\tt CC3} and {\tt CC11} processes are shown
in figure~\ref{frat311} without and in figure~\ref{frat311i} with QED
corrections.
Quantitatively, the figures again do not differ too much.
At the $W$ pair production  peak, the background influence is
smallest, tending to negative contributions of few percents at large
energies.
Below the $W$ pair production threshold, the {\tt CC3} process becomes
heavily suppressed and background becomes equally important if not
dominant.

%
%

%
%

%
%

%
%


\subsection{Conclusions}
%
We determined the so-called background contributions to the off shell
$W$ pair production cross section for the simplest final state
configuration.
Exploiting the inherent symmetries, we arrived at a remarkably compact
presentation for the double differential distribution in the
invariant masses of the $W$ bosons.
With no doubt, the angular distribution in the $W$ production angle is
also quite compact, as our expressions for the {\tt CC3} case
indicate.
Other angular distributions might be accessible as well;
for the neutral current case, there are encouraging indications for
that~\cite{arnd.angular}.
Staying more differential could be welcome to colleagues who search
for anomalous gauge boson couplings, while, on the other hand, the
advantages of the semi-analytical approach are partly lost.
Anyhow, we see no principal problems arising when anomalous couplings
are included.
A step in this direction has been done for the  {\tt CC3}
process~\cite{ancc3}.

Finally, we would like to comment on the prospects of applications of
our techniques to other four-fermion final state topologies.
For the production of four untagged hadronic jets,
one has to add incoherently the $qqgg$ final state, which should be
straightforward.
More involved is the treatment of production of identical
particles and/or of processes with gauge boson exchanges in the $t$
channel.
The first problem is due to permutations of final fermions
in some of the diagrams, the second one by angular dependent
gauge boson propagators, which occurs in e.g. the {\tt CC20} process
and is related to gauge invariance violation~\cite{gauge}.

Time will show to what extent the technical obstacles mentioned will
be overcome in the semi-analytical approach advocated here.

%
\section*{Acknowledgments}
We would like to thank the MPI Munich for the kind hospitality
extended to us in 1992 when this project was started.
Together with M.~Bilenky, we derived analytical results and, also with
the aid of A.~Olchevski, numerically reliable Fortran codes for the
process  {\tt CC11}~\cite{wwteup,LCws}.
At that time, the expressions for the kinematical functions $\cal G$
filled several pages.
In 1994, A.~Leike discovered nice symmetries for the
process {\tt NC24}.
Guided by this, we found extreme simplifications for the process
{\tt CC11}.
M.B. left the project before this level of understanding was reached.
We acknowledge the contributions he made while he was joining us.
Numerous numerical
comparisons with results from Monte Carlo and deterministic approaches
were very helpful.
These were undertaken with A.~Ballestrero, F.~Berends, G.~Montagna,
O.~Nicrosini, R.~Pittau, G.~Passarino, and F.~Piccinini.
Further, we would like to thank B.~Kniehl, H.~K\"uhn,
D.~Lehner, A.~Leike, A. Olchevski, T.~Sjostrand for discussions and
valuable hints.
D.B. is grateful to the Theoretical Physics Division of CERN and
T.R. to the Aspen Center for Physics for the kind hospitality.
\clearpage
\appendix
\def\theequation{\Alph{section}.\arabic{equation}}
\section{Some properties of the auxiliary functions
$\lambda(s,s_1,s_2)$ and ${\cal L}(s;s_1,s_2)$
\label{lamlog}
}
\ezero
Some of the properties of the two auxiliary functions $\lambda$
and $\cal L$ are collected here.
We begin with the $\lambda$ function:
\ba
\lambda &\equiv&
\lambda(s,s_1,s_2)
=
\lambda(s_1,s,s_2)
=
\lambda(s_2,s_1,s)
\nl
&=&
s^2 + s_1^2 + s_2^2 - 2ss_1-2s_1s_2 -2s_2s
\nl
&=&
\left[ s-\left(\sqrt{s_1}+\sqrt{s_2}\right)^2\right]
\left[ s-\left(\sqrt{s_1}-\sqrt{s_2}\right)^2\right].
\label{lam0}
\ea
The $\lambda$  function is symmetric in its arguments.
{}From~(\ref{lam0}), the zeroes may be read off easily.
One is the virtual threshold (and integration boundary),
\ba
\sqrt{s} = \sqrt{s_1} + \sqrt{s_2},
\label{zero}
\ea
the other one is unphysical.
On the mass shell, the following limit holds:
\ba
\sqrt{\lambda(s,M_W^2,M_W^2)}
&=&
s \beta(s,M_W^2) ,
\nl
\beta(s,M_W^2)&=&\sqrt{1-\frac{4M_W^2}{s}}.
\label{beta}
\ea
The logarithm $\cal L$ is defined as follows:
\ba
{\cal L}(s;s_1,s_2)
&=&
{\cal L}(s;s_2,s_1)
\nl
&=&
\frac{1}{\sqrt{\lambda}}
\ln  \frac
{s-s_1-s_2+\sqrt{\lambda}}
{s-s_1-s_2-\sqrt{\lambda}}.
\label{lamb1}
\ea
This function is symmetric in the last two arguments.
For vanishingly small invariant masses, it is:
\ba
\lim_{s_1,s_2\rightarrow 0} {\cal L}(s;s_1,s_2)
&=&
\frac{1}{s}\ln  \frac {s^2}{s_1s_2}.
\label{lamb4}
\ea
The other important limit is approached when $\lambda$ vanishes, which
corresponds to the virtual threshold kinematics:
\ba
\lim_{\lambda\rightarrow 0}
{\cal L}(s;s_1,s_2)
&=&
\frac{2}{s-s_1-s_2} + \ldots
\label{lamb5}
\ea
In the on mass shell limit, the following simplification holds:
\ba
{\cal L}(s;M_W^2,M_W^2)
&=&
\frac{2}{s\beta}
\ln  \frac
{1+\beta}
{1-\beta}.
\label{lamb2}
\ea
In the ultra-relativistic limit, i.e. when
$M_W^2$ becomes vanishingly small compared to $s$, one better uses
\ba
\lim_{M_W^2/s\rightarrow 0}
{\cal L}(s;M_W^2,M_W^2)
&=&
\frac{2}{s} \ln  \frac {s}{M_W^2}.
\label{lamb3}
\ea

\section{Four-momenta and phase space variables}
\label{fmpsv}
\ezero
In this appendix, we express the four-momenta $k_i$ and $p_j$ in terms
of the variables, which are used for the parameterization of the phase
space: $s_1,s_2, \theta, \theta_i, \phi_i$.
Then,
the phase space integration over the squared matrix elements, which
are expressions in the scalar products $k_ik_j, k_ip_l,p_lp_n$,
$i,j=1,2$, $l,n=1,\ldots 4$, may be easily performed.

In the center of
mass frame, 
the initial state vectors $k_1$~and $k_2$ are given
by
\ba
  k_1 & = & \left( k_0,~-k\cos\theta,~0,~k\cos \theta \right) ,
\\
  k_2 & = & \left( k_0,~k \cos\theta,~0,~-k \cos\theta \right),
\ea
with
\ba
  k_0 & = & \frac{\sqrt{s}}{2},
\\
  \left|{\vec k_i}\right| \equiv k  & = &
  \frac{\sqrt{\lambda(s,m_e^2,m_e^2)}}{2\sqrt{s}}.
\ea
This allows to calculate $k_1k_2$.

The products $p_1p_2$ and $p_3p_4$ may be determined in the rest systems
of the corresponding fermion pairs.
For the `compound' $V_1$ (corresponding to the
$W^-$ boson in case of resonant production), the rest system $R$ is
defined by the condition
\ba
p_{12} &=& p_1+p_2,
\\
{\vec p}_{12}^R &=& 0.
\label{r1}
\ea
It is
\ba
p_{1,0}^R &=& \frac{s_1+m_1^2-m_2^2}{2\sqrt{s_1}},
\\
p_{2,0}^R &=& \frac{s_1-m_1^2+m_2^2}{2\sqrt{s_1}}.
\label{p120}
\ea
The 3-momenta of the two fermions are pointing into opposite
directions and we have:
\ba
|{\vec p}_{1}^R| = |{\vec p}_{2}^R| \equiv p_{12}^R
&=& \frac{\sqrt{\lambda(s_1,m_1^2,m_2^2)}}{2\sqrt{s_1}}.
\label{rr1}
\ea
The components of the fermion momenta $p_1$ and $p_2$ in the rest
system
of the fermion pair are
($ p_i^R = \{p_{i,0}^R,p_{i,x}^R,p_{i,y}^R,p_{i,z}^R\}$, i=1,2):
\ba
p_1^R
&=&
\left\{
p_{1,0}^R,
p_{12}^R\sin\theta_1\cos\phi_1,
p_{12}^R\sin\theta_1\sin\phi_1,
p_{12}^R\cos\theta_1
\right\},
\\
p_2^R &=& \left\{
p_{2,0}^R,
-p_{12}^R\sin\theta_1\cos\phi_1,
-p_{12}^R\sin\theta_1\sin\phi_1,
-p_{12}^R\cos\theta_1
\right\} .
\label{p1p2r}
\ea
These relations allow the calculation of the product $p_1p_2$ in terms of the
integration variables; similar relations hold for the other fermion
pair and the product $p_3p_4$.

The remaining Lorentz invariant scalar products $k_ip_j, p_1p_3$, etc.,
may be calculated by boosting
all the final state momenta into the center of mass system.
We will assume that the `compound' $V_1$ moves along the positive $z$ axis.
Then, it is in the center of mass system ($p_{34}=p_3+p_4$):
\ba
p_{12,0} &=& \frac{s+s_1-s_2}{2\sqrt{s}},
\\
p_{34,0} &=& \frac{s-s_1+s_2}{2\sqrt{s}},
\\
|{\vec p}_{12}| = |{\vec p}_{34}|
&=& \frac{\sqrt{\lambda}}{2\sqrt{s}}.
\label{rrr1}
\ea

With the aid of the above definitions the Lorentz transformation from
the center of mass system (and back) may be expressed.
The components of $p_1$ transform from the rest system of $V_1$ into
the center of mass system according to:
\ba
p_{1,0} &=& \gamma_{12}^0 \, p_{1,0}^R + \gamma_{12} \, p_{1,z}^R,
\\
p_{1,x} &=& p_{1,x}^R,
\\
p_{1,y} &=& p_{1,y}^R,
\\
p_{1,z} &=& \gamma_{12}\,  p_{1,0}^R + \gamma_{12}^0 \, p_{1,z}^R,
\ea
where
\ba
\gamma_{12}^0 &\equiv& \frac{p_{12,0}}{\sqrt{s_1}} =
\frac{s+s_1-s_2}{2\sqrt{ss_1}},
\\
\gamma_{12} &\equiv& \frac{|{\vec p}_{12}|}{\sqrt{s_1}} =
\frac{\sqrt{\lambda}}{2\sqrt{ss_1}},
\label{tr1}
\ea
and analogously for $p_2$ and for the other fermion pair.
One should have in mind that
for the `compound' $V_2$ the sign  of the transformation  of the $z$ component
is negative.

As a result of all these relations, we may finally express the components of
the four final state momenta in the center of mass system,
$p_i = \{p_{i,0}, p_{i,x},p_{i,y},p_{i,z}\}$, $i=1,\ldots 4$:
\vspace{.5cm} \\
$\begin{array}{lll}
  p_1 & \!\!\! = & \!\!\!\!
    \left\{ \gamma_{12}^0\!\:p_{1,0}^{R} +
                           \gamma_{12}\!\:p_{12}^R\cos\theta_{1},\:
           p_{12}^R\!\:\sin\theta_{1} \cos\phi_{1},\:
           p_{12}^R\!\:\sin\theta_{1} \sin\phi_{1},\:
           \gamma_{12}^0\!\:p_{12}^R \cos\theta_{1} +
                           \gamma_{12}\!\:p_{1,0}^{R}
            \right\} ,
\\ \\
  p_2 & \!\!\! = & \!\!\!\!
    \left\{ \gamma_{12}^0\!\:p_{2,0}^{R} -
                           \gamma_{12}\!\:p_{12}^R\cos\theta_{1},\;
          -p_{12}^R\!\:\sin\theta_{1} \cos\phi_{1},\:
          -p_{12}^R\!\:\sin\theta_{1} \sin\phi_{1},\:
          -\gamma_{12}^0\!\:p_{12}^R \cos\theta_{1} +
                           \gamma_{12}\!\:p_{2,0}^{R}
            \right\} ,
\\ \\

  p_3 & \!\!\! = & \!\!\!\!
    \left\{ \gamma_{34}^0\!\:p_{3,0}^{R} -
                           \gamma_{34}\!\:p_{34}^R \cos\theta_{2},\:
           p_{34}^R\!\:\sin\theta_{2} \cos\phi_{2},\:
           p_{34}^R\!\:\sin\theta_{2} \sin\phi_{2},\:
           \gamma_{34}^0\!\:p_{34}^R \cos\theta_{2} -
                           \gamma_{34}\!\:p_{3,0}^{R}
            \right\} ,
\\ \\
  p_4 & \!\!\! = & \!\!\!\!
    \left\{ \gamma_{34}^0\!\:p_{4,0}^{R} +
                           \gamma_{34}\!\:p_{34}^R\cos\theta_{2},\;
          -p_{34}^R\!\:\sin\theta_{2} \cos\phi_{2},\:
          -p_{34}^R\!\:\sin\theta_{2} \sin\phi_{2},\:
          -\gamma_{34}^0\!\:p_{34}^R \cos\theta_{2} -
                           \gamma_{34}\!\:p_{4,0}^{R}
            \right\} .
\end{array}$
\ba
  \label{pvect}
\ea
With these relations, any of the scalar products may be
expressed in terms of integration variables.
It may be seen explicitly that in the center of mass system the
$(p_1+p_2)$ and $(p_3+p_4)$,
as well as  $(k_1+k_2)$, are independent of all angular variables
while  $(p_1-p_2)$ and  $(p_3-p_4)$, as well as $(k_1-k_2)$, depend
each on only one angle $\theta_1$, $\theta_2$, $\theta$, respectively.

%
\section{Phase space integrals
\label{psi}}
\ezero
%
The gauge boson propagators depend on one of the invariant masses
squared $s,s_1,s_2$ and are left for numerical integrations.
Thus, the only angular integrals will arise from fermion
propagators.
In this respect, the {\tt CC3} process is distinguished by its
simplicity since the final state integrations factorize completely and
may be performed by tensor integration  over $\theta_i,\phi_i$
and then the neutrino propagator in
the {\tt crab} diagram contains the only nontrivial angular dependence:
\ba
D_{\nu}
&\sim& \frac{4}{
\left[ (k_1-k_2) - (p_1+p_2) + (p_3+p_4)\right]^2
}
= \frac{2}{s-s_1-s_2-\sqrt{\lambda}\cos\theta}
\equiv \frac{1}{t_{\nu}}.
\label{npr}
\ea
Here, only the difference $(k_1-k_2)$ depends on an angle  (see
the final remarks in appendix~\ref{fmpsv}.
The one dimensional integration is:
\ba
\left[ A \right]_{\theta} \equiv \frac{1}{2}
  \int\limits_{-1}^{+1} d\cos\theta~A.
\ea
Besides $[\cos^n\theta]_{\theta}=(1,0,1/3)$ for $n=0,1,2$,
the following integrals are used:
\ba
\left[ \frac{1}{\,t_{\nu}\,} \right]_{\theta}  & = &
{\cal L} (s;s_1,s_2),
\\
\left[ \frac{1}{\,t_{\nu}^2\,} \right]_{\theta} & = &
        \frac{1}{s_1 s_2}.
\label{table_t}
\ea

The background contributions contain either one or two (equal or
different) fermion propagators.
An example is:
\ba
D_{f_1} &\sim& \frac{4}{\left[ (k_1+k_2) - (p_1-p_2) + (p_3+p_4)\right]^2}
=  \frac{2}{s_1-s_2-s+\sqrt{\lambda}\cos\theta_1}
\equiv  \frac{1}{t_{f_1}}.
\label{fpr}
\ea
which occurs in the second of the Feynman diagrams of figure~\ref{fig.2}.
Here, only $(p_1-p_2)$ depends on an angle.
The other propagators are expressed similarly.
Evidently, the angular integrations over $\phi_1,\phi_2$ are not
influenced and those over $\theta, \theta_1, \theta_2$ factorize and
may be performed independently using the above table of integrals,
thereby permuting $s,s_1,s_2$ in the answers as needed.
After a careful book-keeping, the background contributions are not
much more involved than the {\tt CC3} case.
%
\section{The neutral current kinematical functions
\label{gnc224}}
\ezero
The off shell $Z$ pair production may be described by one kinematical
function~\cite{nc2}:
\ba
{\cal G}_{\tt NC2}(s;s_1,s_2)
&=&
\frac
{s^2+\left(s_1+s_2\right)^2}
{s-s_1-s_2}
{\cal L}(s,s_1,s_2) - 2
{}.
\label{gnc2}
\ea

The process {\tt NC24}, which proceeds via off shell gauge boson
pairs, but also via single resonant diagrams of the {\tt deer} type,
may be described by two functions ${\cal G}_{\tt NC2}$ and
${\cal G}_{\tt NC24}$.
The latter is~\cite{nc24}:
\ba
 {\cal G}_{\tt NC24}(s;s_1,s_2) &=&
s s_1 s_2 \times
 \frac{3}{\lambda^2}\Biggl\{
 {\cal L}(s_2;s,s_1) {\cal L}(s_1;s_2,s)
          4s\Bigl[ss_1(s-s_1)^2+ss_2(s-s_2)^2
\nl
&&\hspace{4.5cm}+~s_1s_2(s_1-s_2)^2\Bigr]
\nl
& & +~(s+s_1+s_2)\Biggl[
            {\cal L}(s_2;s,s_1)
          2s\left[(s-s_2)^2+s_1(s-2s_1+s_2)\right] \nl
& & \hspace{2.5cm} +~ {\cal L}(s_1;s_2,s)
          2s\left[(s-s_1)^2+s_2(s+s_1-2s_2)\right] \nl
& & \hspace{4.9cm} +~ 5s^2-4s(s_1+s_2)-(s_1-s_2)^2 \Biggr] \Biggr\}.
\label{gnc24}
\ea
The function ${\cal G}_{\tt NC24}$ may be identified in the integrand
on the right hand side of equation~(2.11) in~\cite{nc24kk}.
The quantity $I_4$, which is defined there and contains higher order
axial corrections to the $Z$ width, is related to ${\cal G}_{\tt
  NC24}$ as follows:
\ba
I_4 =  \int \frac{ds_1ds_2}{s_1s_2}
\frac{\sqrt{\lambda}}{3s}
{\cal   G}_{\tt NC24}(s;s_1,s_2) = \frac{\pi^2}{3}-\frac{15}{4}.
\ea

For numerical applications, the following limit is useful:
\ba
\lim_{\lambda\rightarrow 0}
{\cal G}_{\tt NC24}(s;s_1,s_2)
&=&
\frac{1}{s_1s_2} \left[
- \frac{9s^2-\Delta^2}{s^2-\Delta^2} - \frac{2}{3}
\frac{\lambda}{(s^2-\Delta^2)^3}\left(9s^4+
6s^2\Delta^2+\Delta^4\right) + {\cal O}(\lambda^2) \right] ,
\ea
with $\Delta = s_1-s_2$.


\newpage
\begin{thebibliography}{99}
\label{bibl}
\bibitem{lepproc}
G. Altarelli (ed.),
Proc. of the Workshop on Physics at LEP~2, CERN 1995, CERN Yellow
Report in preparation.
\bibitem{lcproc}
P. Zerwas (ed.),
Proc. of the European Workshop on Physics with $e^+e^-$ Linear
Colliders, Annecy, Gran Sasso, Hamburg, 1995, in preparation.
\bibitem{excalibur}
F.A. Berends, R.~Kleiss and R. Pittau,
Nucl. Phys. {\bf B424} (1994) 308; 
Nucl. Phys. {\bf B426} (1994) 344; 
Fortran program {\tt EXCALIBUR}, {\it Comput. Phys. Commun.} {\bf 85}
(1995) 437;
\\
R. Pittau,
{\it Phys. Letters} {\bf B335} (1994) 490.
\bibitem{wwteup}
D.~Bardin, M.~Bi\-len\-ky, D.~Leh\-ner, A.~Ol\-chev\-ski
and T.~Riemann,
in:
T.~Riemann and J.~Bl\"umlein (eds.),
Proc. of the Zeuthen Workshop on Elementary Particle Theory --
Physics at LEP200 and Beyond, Teupitz, Germany,
April 10--15, 1994,
{\it Nucl. Phys.}  (Proc. Suppl.) {\bf 37B} (1994) p. 148.
\bibitem{nc24}
D.~Bardin, A.~Leike and T.~Riemann,
{\it Phys. Letters} {\bf B344} (1995) 383.
\bibitem{nc24h}
D.~Bardin, A.~Leike and T.~Riemann,
{\it Phys. Letters} {\bf B353} (1995) 513.
\bibitem{comphep}
E. Boos et al.,
{\tt CompHEP}: computer system for calculation of particle collision
characteristics at high energies, version 2.3 (1991),
Moscow State Univ. preprint MGU-89-63/140 (1989);
preprint KEK 92-47 (1992).
\bibitem{form}
J.A.M.~Vermaseren, {\it Symbolic Manipulation with {\tt FORM}},
Computer Algebra Nederland, Amsterdam, 1991.
\bibitem{muta}
T.~Muta, R.~Najima and S.~Wakaizumi,
{\it Mod. Phys. Letters} {\bf A1} (1986) 203.
\bibitem{nc2}
V. Baier, V. Fadin and V. Khoze, {\it Sov. Phys. JETP} {\bf 23}
(1966) 104;
\\
R.W.~Brown and K.O.~Mikaelian, {\it Phys. Rev.} {\bf D19} (1979) 922;
%
\\
M.~Cvetic and P.~Langacker,
{\it Phys. Rev.} {\bf D46} (1992) 4943; E: {\bf D48} (1993) 4484.
\bibitem{pythia}
T. Sjostrand, Fortran programs {\tt PYTHIA~5.7} and {\tt JETSET~7.4}
(Oct. 1994), {\it Comput. Phys. Commun.} {\bf 82} (1994) 74;
updated version: Lund preprint LU TP 95--20 (1995) [hep-ph/9508391].
\bibitem{wpairs}
D. Bardin et al.,
Report of the working groups on $W$ pair production and Monte Carlo
generators, in~\cite{lepproc}.
\bibitem{anom}
G.~Gounaris et al.,
Report of the working group on anomalous gauge boson couplings,
in~\cite{lepproc}.
\bibitem{vanO}
A.~Aeppli, preprint BNL-46819 (1991);
\\
A.~Aeppli and D.~Wyler,
{\it Phys. Letters} {\bf B262} (1991) 125;
\\
A. Aeppli, G.J. van Oldenborgh and D. Wyler,
preprint PSI--PR--93--22;
\\
G.J. van Oldenborgh, P.J.~Franzini and A. Borrelli,
{\it Comp. Phys. Commun.} {\bf 83} (1994) 14.
\bibitem{berends}
F.A.~Berends, G.~Burgers and W.L.~van Neerven,
{\it Nucl. Phys.} {\bf B329} (1990) 429; E: {\bf B304} (1998) 921.
\bibitem{kuehn}
B.A. Kniehl, M. Krawczyk, J.H. K\"uhn and R.~Stuart,
{\it Phys. Letters} {\bf B209} (1988) 337.
\bibitem{wwqed}
D. Bardin, M. Bilenky, A.~Olchevski and T.~Riemann,
{\it Phys. Letters} {\bf B308} (1993) 403;
revised version: hep-ph/9507277, E: to appear.
\bibitem{zzqed}
D. Bardin, D. Lehner and T.~Riemann,
in:
Proc. of the IX Int. Workshop on High Energy Physics and Quantum Field
Theory, Zvenigorod, Russia, Sept 1994, in press;
preprint DESY 94--216 (1994) [hep-ph/9411321].
\bibitem{coulomb}
D. Bardin, W. Beenakker and A.~Denner,
{\it Phys. Letters} {\bf B317} (1993) 213;
\\
see also:
V.S.~Fadin, V.A.~Khoze and A.D.~Martin,
{\it Phys. Letters} {\bf B311} (1993) 311;
 {\bf B320} (1994)141; {\it Phys. Rev.} {\bf D49} (1994) 2247.
\bibitem{kufa}
E.A. Kuraev and V.S. Fadin, {\it Sov. J. Nucl. Phys.} {\bf 41} (1985) 466.
\bibitem{gentle}
D.~Bardin and T.~Riemann,
Fortran program {\tt GENTLE}, version {\tt CC11} (June 1995).
\\
Version {\tt CC11} originates from:
D. Bardin, M. Bilenky, A.~Olchevski and T.~Riemann,
Fortran program {\tt GENTLE}:
A semi-Monte Carlo {\tt GEN}erator of the radiative {\tt T}ail for
{\tt LE}P~200.
\bibitem{zdecay}
A.~Akhundov, D.~Bardin and T.~Riemann,
{\it Nucl. Phys.} {\bf B276} (1986) 1;
\\
 W. Beenakker and W.~Hollik, {\it Z. Physik} {\bf C40} (1988) 141;
\\
J. Bernabeu, A. Pich and A. Santamaria,
{\it Phys. Letters} {\bf B200} (1988) 569;
\\
see also: D. Bardin, W. Hollik and G. Passarino (eds.),
Reports of the Working Group on Precision Calculations for the $Z$ Resonance,
CERN Yellow Report CERN 95--03 (1995).
\bibitem{wdecay}
D.~Bardin, S.~Riemann and T.~Riemann,
{\it Z. Physik} {\bf C32} (1986) 121;
\\
F.~Jegerlehner, {\it Z. Physik} {\bf C32} (1986) 425; E: {\bf 38}
(1988) 519;
\\ A.~Denner and T.~Sack,
{\it Z. Physik} {\bf C46} (1990) 653.
\bibitem{wphact}
E.~Accomando and A.~Ballestrero, Fortran program {\tt WPHACT}.
\bibitem{wwgenpv}
G.~Montagna, O.~Nicrosini and F.~Piccinini, Fortran program {\tt
  WWGENPV}, A Monte Carlo Event Generator for Four-Fermion Production
in $e^+ e^- \to W^+ W^- \to 4f$, preprint CERN-TH/95--100 (1995), to
appear in {\it Comp. Physics Commun}.
\bibitem{wto}
G. Passarino, Fortran program {\tt WTO}.
\bibitem{arnd.angular}
A. Leike,
Contribution to the working group on Higgs boson physics of the
Workshop~\cite{lcproc}.
\bibitem{ancc3}
F.~Berends and A.I.~van~Singhem,
Leiden preprint INLO--PUB--7/95 [hep-ph/9506391];
see also~\cite{anom}.
\bibitem{gauge}
E.N. ~Argyres et al., preprint INLO--PUB--8/95 [hep-ph/9507216].
\bibitem{LCws}
D. Bardin, M. Bilenky, A.~Olchevski and T.~Riemann,
preprint CERN-TH. 7102/93 (1993),
in:
P.M.~Zerwas (ed.),
Proc. of the Workshop on $e^+e^-$ Collisions at 500 GeV: The Physics
Potential, Munich, Annecy, Hamburg (Nov 1992 to April 1993),
DESY 93-123C (1993) p.~159.
\bibitem{nc24kk}
B.A.~Kniehl and J.H.~K\"uhn, {\it Nucl. Phys.} {\bf B329} (1990) 547.
\end{thebibliography}
\end{document}